\documentclass[final]{siamltex}

\usepackage{epsf,xspace,graphicx}
\usepackage{hyperref}
\usepackage{breakurl}
\newcommand{\bds}[1]{\boldsymbol{#1}}

\usepackage{amsfonts,amsmath,graphicx,epsfig}
\usepackage{subfigure,graphics,amssymb,amsxtra,color}
\usepackage[ruled,vlined]{algorithm2e}
\usepackage[ansinew]{inputenc}
\usepackage{nicefrac}

\newtheorem{remark}{Remark} 

\newcommand{\VH}{\mathbf{V_H}}

\newcommand{\SW}{\bf \Sigma_w}
\newcommand{\SWi}{{\mathbf{\Sigma}_{\bf w}^{-1}}}

\newcommand{\UW}{{\mathbf{U}_{\bf w}}}
\newcommand{\UWt}{{\mathbf{U}_{\bf w}^\dagger}}
\newcommand{\LW}{\mathbf{\Lambda}_\mathbf{w}}
\newcommand{\LWi}{\mathbf{\Lambda}_\mathbf{w}^{-1}}
\newcommand{\LWisqrt}{\mathbf{\Lambda}_\mathbf{w}^{-\nicefrac{1}{2}}}

\newcommand{\M}{\mathbf{M}}
\newcommand{\Ms}{\mathbf{M}^\star}
\newcommand{\Mst}{\mathbf{M}^{\star\dagger}}
\newcommand{\Mt}{\mathbf{M}^\dagger}
\newcommand{\UM}{\mathbf{U_M}}
\newcommand{\UMs}{\mathbf{U}^\star_\mathbf{M}}
\newcommand{\UMst}{{\mathbf{U}_\mathbf{M}^{\star\dagger}}}

\newcommand{\LM}{\mathbf{\Lambda}_\mathbf{M}}

\newcommand{\LMq}{{\mathbf{\Lambda}_\mathbf{Q}}}
\newcommand{\LMqs}{{\mathbf{\Lambda}_\mathbf{Q}^{\star }}}
\newcommand{\LMs}{{\mathbf{\Lambda}_\mathbf{M}^\star}}
\newcommand{\LMst}{{\mathbf{\Lambda}_\mathbf{M}^{\star\dagger}}}
\newcommand{\LMt}{{\mathbf{\Lambda}_\mathbf{M}^\dagger}}
\newcommand{\VM}{\mathbf{V_M}}
\newcommand{\VMs}{\mathbf{V}^\star_\mathbf{M}}
\newcommand{\VMst}{\mathbf{V}^{\star\dagger}_\mathbf{M}}
\newcommand{\VMt}{\mathbf{V}^{\dagger}_\mathbf{M}}

\newcommand{\E}{\mathbf{E}}  
\newcommand{\Es}{{\bf E}^\star} 
\newcommand{\UE}{{\bf U}_{\bf E}}  
\newcommand{\UEs}{{\bf U}_{\bf E}^\star} 
\newcommand{\UEst}{{{\bf U}_{\bf E}^{\star\dagger}}} 
\newcommand{\UEt}{{{\bf U}_{\bf E}^{\dagger}}} 
\newcommand{\LE}{{\mathbf{\Lambda}_\mathbf{E}}}

\newcommand{\UX}{{\bf U}_{\bf x}}  
\newcommand{\UXt}{{{\bf U}_{\bf x}^\dagger}}  
\newcommand{\SX}{{\bf \Sigma_\x}}

\newcommand{\LX}{{\mathbf{\Lambda}_\mathbf{x}}}

\newcommand{\Os}{{{\bf \Pi}^\star}}
\newcommand{\Ost}{{{\bf \Pi}^{\star\dagger}}}

\newcommand{\U}{\mathbf{U}}  

\newcommand{\Ixy}{\mathcal{I}\left({\bf x};{\bf y}\right)}

\newcommand{\Hx}{\mathcal{H}_x\left(\x\right)}

\newcommand{\PP}{\mathbf{P}}
\newcommand{\PPt}{\mathbf{P}^\dagger}

\newcommand{\A}{\mathbf{A}}

\newcommand{\I}{\mathbf{I}}
\newcommand{\J}{\mathbf{J}}
\newcommand{\X}{\mathbf{X}}
\newcommand{\x}{\mathbf{x}}
\newcommand{\y}{\mathbf{y}}

\newcommand{\n}{\mathbf{n}}
\newcommand{\w}{\mathbf{w}}

\newcommand{\tr}{{\sf tr}}

\title{Communications-Inspired Projection Design with Application to Compressive Sensing}
\author{William~R.~Carson\footnotemark[2]\ \footnotemark[4]
\and Minhua~Chen\footnotemark[3]
\and Miguel~R.~D.~Rodrigues\footnotemark[2]\ \footnotemark[5]
\and Robert~Calderbank\footnotemark[3]
\and Lawrence~Carin\footnotemark[3]}

\begin{document}
\maketitle

\renewcommand{\thefootnote}{\fnsymbol{footnote}}

\footnotetext[2]{W. R. Carson and M. R. D. Rodrigues were supported by the Funda{\c c}{\~ a}o para a Ci{\^e}ncia e a Tecnologia through the research project PTDC/EEA-TEL/100854/2008.}
\footnotetext[3]{M. Chen, R. Calderbank and L. Carin are with Department of Electrical Engineering, Duke University, USA. Their work was supported in part by NSF under Grant DMS-0914892, by ONR under Grant N00014-08-1-1110, and by DARPA under grant N66001-11-1-4002 as part of the KeCom Program.}
\footnotetext[4]{W. R. Carson is now with PA Consulting Group, Cambridge Technology Centre, Melbourn, UK.}
\footnotetext[5]{M. R. D. Rodrigues is now with Dept. E \& EE, University College London, London, U.K.}
\footnotetext{ A part of the material presented in this work has been published in IEEE ICASSP 2012.}
\renewcommand{\thefootnote}{\arabic{footnote}}

\begin{abstract}

We consider the recovery of an underlying signal $\x\in\mathbb{C}^m$ based on projection measurements of the form $\y=\M\x+\w$, where $\y\in\mathbb{C}^\ell$ and $\w$ is measurement noise; we are interested in the case $\ell\ll m$. It is assumed that the signal model $p(\x)$ is known, and $\w\sim\mathcal{CN}(\w;\bds{0},\SW)$, for known $\SW$. The objective is to design a projection matrix $\M\in\mathbb{C}^{\ell\times m}$ to maximize key information-theoretic quantities with operational significance, including the mutual information between the signal and the projections $\mathcal{I}(\x;\y)$ or the R{\' e}nyi entropy of the projections $\mbox{ h}_\alpha \left( \y \right)$ (Shannon entropy is a special case). By capitalizing on explicit characterizations of the gradients of the information measures with respect to the projections matrix, where we also partially extend the well-known results of Palomar and Verd{\'u} from the mutual information to the R{\' e}nyi entropy domain, we unveil the key operations carried out by the optimal projections designs: \emph{mode exposure} and {mode alignment}. Experiments are considered for the case of compressive sensing (CS) applied to imagery. In this context, we provide a demonstration of the performance improvement possible through the application of the novel projection designs in relation to conventional ones, as well as justification for a fast online projections design method with which state-of-the-art adaptive CS signal recovery is achieved.

\end{abstract}



\pagestyle{myheadings}
\thispagestyle{plain}
\markboth{W. R. Carson, M. Chen, M.R.D. Rodrigues, R. Calderbank and L. Carin}{Communications-Inspired Projection Design}

\section{Introduction} \label{introduction}

C{ompressive} sensing (CS) \cite{Can06,Donoho2006} has recently emerged as an important area of research in image sensing and processing. Compressive sensing has been particularly successful in multidimensional imaging applications, including magnetic resonance \cite{Lustig2007}, spectral imaging \cite{Gehm07,Sun11} and video \cite{Shankar,CAVE_0311}. Conventional sensing systems typically first acquire data in an uncompressed form ($e.g.$, individual pixels in an image) and then perform compression subsequently, for storage or communication. In contrast, CS involves acquisition of the data in an already compressed form, reducing the quantity of data that need be measured in the first place. In CS the underlying signal to be measured is projected onto a set of vectors \cite{Candes2006,Donoho2006}, and one must perform an inverse problem to recover the underlying signal of interest. 

There are two hallmarks of the original CS theory. First, the projection vectors were usually constituted uniformly at random. Second, the underlying signal model used to regularize the inverse problem was based on the assumption that the underlying signal could be sparsely represented in terms of an orthonormal basis or frame. However, even in some of the early CS studies, it was recognized that improved performance could be achieved with projection vectors designed to the underlying signal of interest \cite{BCS,Optics,Neifeld}, rather than using random projections. Further, it has recently been recognized that a signal model based upon sparsity is often overly primitive, and model-based CS \cite{Baraniuk10}, wherein improved signal models are employed, may yield improved CS performance (high-quality signal recovery with fewer projection measurements).  Signal models that have been considered include the Gaussian mixture model (GMM) \cite{Chen10}, union-of-subspace models \cite{Yonina_block}, and manifold models \cite{Wakin}. 

In this paper our goal is to design CS projection matrices (``measurement kernels'') matched to a general statistical signal model. Specifically, if the underlying signal to be measured is $\x\in\mathbb{C}^m$, it is assumed that we have access to a general signal model, represented statistically by density function $p(\x)$. Our objective is to design the projection matrix to maximize the mutual information between the underlying signal and the observed compressive measurements or to maximize the R{\'e}nyi of the compressive measurements.

The key to the approach considered in this paper is the realization that the projection-design problem for CS systems (subject to a power constraint) exhibits parallels with the precoder design problem for multiple-input--multiple-output (MIMO) communications systems: in the communications problem a source is being matched to a channel whereas in CS a channel, or equivalently the noise covariance, is being matched to the source. This link has also been recognized recently by Schnitter \cite{Schniter11}, who has provided projections designs for sources modelled by multivariate Gaussian distributions, as well as by Carson \emph{et al.} \cite{Carson12}, who have also provided designs for general multivariate source distributions. With the precoder design problem exhibiting a long tradition in the information theory and communications field, this link also provides the means to translate, with appropriate modifications, much of the design know-how and experience from the communications domain to CS.

The traditional problem of precoder design for MIMO Gaussian channels has been drawing on various performance metrics relevant for data communications. Common precoder design approaches aim to maximize the system signal-to-noise ratio (SNR) and the system signal-to-interference-plus-noise ratio (SINR) \cite{Palomar03,Scaglione02} or minimize the system error probability \cite{Bergman08}, \cite{Goparaju11}. Another emerging precoder design approach imbued with operational significance is based on the maximization of the mutual information between the input and the output of the system \cite{Lozano06,Payaro09_hessian,Lamarca09,PerezCruz10,Xiao11}. This novel design principle has been shown to yield considerable rate gains in a variety of communications scenarios, due to the fact that, in addition to adapting to the channel characteristics, the designs also adapt to important features necessary to achieve high-rate reliable communications (the designs conform to the exact characteristics rather than only to the second-order statistics of the signaling scheme, as in traditional approaches (see \cite{Palomar03}, \cite{Scaglione02})). The basis of the emergence of the mutual information based designs have been fundamental connections between information theory and estimation theory, which have unveiled the interplay between mutual information and the minimum mean-squared error (MMSE) in scalar Gaussian channels \cite{Guo05} or mutual information and the MMSE matrix in vector Gaussian channels \cite{Palomar06}. These results offer a means to bypass the absence of closed-form mutual information expressions for MIMO Gaussian channels driven by arbitrary (non-Gaussian) signaling schemes.

The operational significance of mutual information, which acts as the rationale for its use as the basis of a plethora of designs, is well known not only in data communications -- it represents the highest reliable information transmission rate in a single-user channel driven by a specific signalling scheme -- but also in other domains. For example, in classification problems mutual information relates (through bounds) to the Bayesian error probability of the classifier \cite{Hellman70}; and, in regression problems mutual information relates (through bounds) to the reconstruction error \cite{Prasad10}. 

We consider design of the measurement kernel based upon maximizing the mutual-information between the underlying signal $\x$ and the compressive measurement $\y$. We also consider design based upon maximizing the R{\'e}nyi entropy of $\y$, where the latter represents a generalization with operational relevance \cite{Erdogmus04}. The projection design will be implemented in practice using gradient descent, and we demonstrate that for a GMM signal model the gradient of R{\'e}nyi entropy with respect to the design matrix may be expressed analytically, for a special parameter setting. Further, we recover the gradient of Shannon entropy as a special case of the R{\'e}nyi result.

The article considers both theoretical results, which disclose key operations effected by the projection designs, as well as experimental results that demonstrate the merit of the approach as applied to a practical CS imaging problem. One key operation relates to the notion of \emph{mode alignment} in mutual information based designs: the modes of the source, which depending on the source statistical model are given by the eigenvectors of the source covariance matrix or the eigenvectors of the source MMSE matrix, have to align with the modes (eigenvectors) of the noise covariance matrix as a means to improve performance. This role can also be conceptually appreciated by viewing the measurement kernel as a sieve that aligns relevant statistical features of the source to the statistical features of the noise, in order to disclose relevant information for reconstruction. The relevance of mode alignment, which is typically absent in communications problems\footnote{This operation is absent in precoder designs for MIMO Gaussian channels driven by Gaussian inputs, due to the fact that the signal covariance is often taken to be white, but is present in precoder designs for MIMO Gaussian channels driven by non-Gaussian inputs. The role of a certain permutation operation in the precoder design is hinted at by Lamarca in \cite{Lamarca09}.}, has also been recently unveiled in radar applications \cite{Tang10}.\footnote{Note that Schniter \cite{Schniter11} does not recognize the role of mode alignment due to the statistical assumptions about the source and noise covariances: this operation is not present when the source covariance is the identity matrix or when the noise covariance is also the identity matrix.}  Overbridging the theoretical and practical results is also the formal justification of a low-complexity high-performance online projections strategy, the partial direction sensing method (PSD) \cite{Sapiro}, which brings together the main operational features of the optimal measurement kernel designs, including mode alignment.

The detailed contributions of the article include:

\begin{itemize}

\item Recognition that recent advances in communications, which relate to the design of precoders for MIMO communications channels, carry over to CS, leading to a communications-inspired kernel design framework for CS applications.

\item Proposal of mutual information based offline -- where a set of projections is optimized simultaneously -- and online -- where the individual projections are optimized sequentially -- kernel designs.  
The article unveils key operations carried out by the optimal kernel designs for multivariate Gaussian sources and general multivariate sources, including the operations of source and noise modes exposure, mode weighting and mode alignment. Particular emphasis is given to the role of mode alignment as a means to improve further the reconstruction performance in compressive sensing applications.

\item Proposal of R{\'e}nyi entropy based kernel designs. The article also underlines some relations between the mutual information (or Shannon entropy) and the R{\'e}nyi entropy based kernel constructions.

\item Formal rationale for the PDS strategy \cite{Sapiro}, which is based on the operational insight unveiled by the theoretical characterizations of the optimal kernel designs.

\item Partial generalizations of the I-MMSE identity from the mutual information (or Shannon entropy) to the R{\'e}nyi entropy domain.

\item A range of experimental results that illustrate the benefit of the novel measurement kernel designs in relation to the conventional random ones.

\end{itemize}


The remainder of the article is organized as follows. In Section \ref{notation} we briefly summarize the notation used throughout. Section \ref{section_system_model} reviews the modeling and design approach, introducing key system assumptions. Section \ref{sec_mutual_information} introduces the optimal kernel design based on the Shannon-based mutual information metric -- this builds upon work on the communications field on precoder design for MIMO channels driven by Gaussian inputs and arbitrary inputs. Section \ref{sec_Renyi_entropy} introduces the optimal kernel design based on the R{\' e}nyi entropy metric, taking advantage of the closed-form expressions available for a GMM source. Section \ref{sec_results} provides the body of evidence that demonstrates the performance improvement possible through the application of the projections designs put forth in previous sections. We consider examples based on offline kernel design, based upon the prior signal model, as well as online kernel design based upon sequential update of the posterior, all within the context of a GMM signal representation, which yields analytic CS inversion. Section \ref{sec_conclusions} draws the main conclusions. The Appendices contain proofs and supporting mathematical derivations.

%
%
%

\section{Notation and Definitions} \label{notation}

In the following text scalar quantities are denoted by italics, vectors are denoted by boldface lower case letters and matrices are denoted by boldface upper case letters. 
The projection of scalar $x$ onto the non-negative orthant is denoted $(x)^+ \triangleq \max(0, x)$.
The superscript $\bf (\cdot)^{\star}$ is used to denote an optimal solution and the superscripts $(\cdot)^{T}$, $\bf (\cdot)^{*}$ and $\bf (\cdot)^{\dagger}$ denote transpose, conjugate and conjugate transpose operators, respectively. 
The element in the $i$-th row and $j$-th column of the matrix $\X$ is denoted by $[\X]_{i,j}$.
The trace of a matrix is denoted $\tr( \cdot )$. The diagonal matrix with diagonal elements given by either vector $\x$ or the diagonal elements of matrix $\X$ is denoted by ${\sf Diag}( \x)$ or ${\sf Diag}(\X)$, respectively.

We refer frequently to the following special matrices and sets: the $n \times n$ identity matrix is denoted $\I_n$, the $n \times n$ flipped identity matrix with ones on the anti-diagonal is denoted $\J_n$, the $m \times n$ matrix of all zeros is denoted ${\bf 0}_{m \times n}$; the sub-scripts may be dropped where no confusion may arise.
The set of all $n \times n$ unitary matrices is denoted $\mathbb{S}^n$ and the set of $m \times n$ complex matrices is denoted $\mathbb{C}^{m \times n}$.

The notation $\x \sim \mathcal{CN}(\x; \boldsymbol{\mu},{\bf \Sigma})$ denotes a random variable $\x$ which is circularly symmetric complex Gaussian distributed with mean $\boldsymbol{\mu}$ and covariance matrix $\bf \Sigma$.

\section{Modelling and Design Approach}\label{section_system_model}

In CS, we aim to reconstruct the signal of interest $\x \in \mathbb{C}^{m}$ based on a small number of noisy projections:
\begin{equation}
\y = \M~\x + \w, \quad\y\in  \mathbb{C}^{\ell}
\label{eqn_CS_noisy}
\end{equation}
with $\ell \leq m$ and where $\M \in \mathbb{C}^{\ell \times m}$ is the kernel (or projection) matrix and $\w$ represents zero-mean circularly symmetric complex Gaussian noise with positive definite covariance matrix $\SW$, i.e., $\w \sim \mathcal{CN}(\w; {\bf 0},\SW)$. 
The action of the kernel can be understood in terms of two separate projections and a power allocation (or stretching) operation, which are associated with the matrices in its singular value decomposition (SVD) given by:
\begin{align}
\M =\UM ~ \LM~ \VMt
\end{align}
where $\LM= \left[{\sf Diag} \left( \sqrt{\lambda_{M_1}}, \ldots, \sqrt{\lambda_{M_\ell}} \right) \hspace{3mm} {\bf 0}_{\ell \times (m - \ell)}\right] \in \mathbb{R}^{\ell \times m}$, $\UM \in \mathbb{S}^{\ell}$, $\VM \in \mathbb{S}^{m}$, and ${\lambda_M}_1 \geq {\lambda_M}_2 \geq \ldots \geq {\lambda_{M_\ell}} \geq 0$ correspond to the (non-negative) eigenvalues of $\M ~\Mt$.

Both the signal and the noise covariance matrices are positive (semi-)definite and can also be represented in terms of their eigenvalue decomposition (projections and power allocation). 
In particular, the signal covariance matrix is given by:
\begin{align}
\SX &= \UX ~ \LX ~ \UXt
\end{align}
where $\UX \in \mathbb{S}^{m}$, $\LX = {\sf Diag} \left( {\lambda_{x_1}}, \ldots, {\lambda_{x_m}} \right)$ and ${\lambda_{x_1}} \geq {\lambda_{x_2}} \geq \ldots \geq {\lambda_{x_m}} \geq 0$ are the (non-negative) eigenvalues of $\SX$.
Similarly, the noise covariance matrix is given by:
\begin{align}
\SW = \UW ~\LW ~  \UWt
\end{align}
where $\UW \in \mathbb{S}^\ell$, $\LW = {\sf Diag} \left( {\lambda_{w_1}}, \ldots, {\lambda_{w_\ell}} \right)$ and $0 \leq {\lambda_w}_1 \leq {\lambda_w}_2 \leq \ldots \leq {\lambda_w}_\ell$ correspond to the (non-negative) eigenvalues of $\SW$.

Our design approach, which relies not only on a statistical model for the noise but also on the signal, draws on specific quantitative metrics in order to conceive and compare various possible kernel designs. A natural metric, which relates to the best achievable reconstruction error, is the (non-linear) MMSE given by:
\begin{align}
\text{MMSE} =  \mathbb{E} \left\{ \tr \left[\left({\bf x} - \mathbb{E} \left\{{\bf x}|{\bf y}\right\}\right)\left({\bf x} - \mathbb{E} \left\{{\bf x}|{\bf y}\right\}\right)^\dagger  \right] \right\}
\end{align}
that involves the use of conditional mean estimation to recover the signal of interest from the noisy projections, i.e., $\hat{\x}(\y) = \mathbb{E} \left\{{\bf x}|{\bf y}\right\}$ \cite{Kay01}. We, however, capitalize on information-theoretic metrics, most notably the mutual information and R{\'e}nyi entropy based on the fact that mutual information and R{\'e}nyi entropy - in view of recent developments in information theory and communications - appear to be more amenable to mathematical analysis than the non-linear MMSE. In addition, it is also possible to bound the MMSE via the mutual information as follows \cite{Prasad10}: 
\begin{align}
\text{MMSE} \geq \frac{1}{2 \pi e} \exp 2\left[\Hx - \Ixy\right].
\end{align}
where $\Hx$ denotes the differential entropy of $\x$ and $\Ixy$ denotes the mutual information between $\x$ and $\y$.

The crux of our design approach, which we also partially extend from the mutual information to the R{\'e}nyi entropy metric, is a fundamental result that links the gradient with respect to some parameters of the mutual information between the input and the output of a linear vector Gaussian channel model and the MMSE matrix associated with the model: known as the I-MMSE relationship. This result, which was originally put forth for the linear scalar Gaussian model by Guo, Shamai and Verd{\' u} \cite{Guo05} and later for linear vector Gaussian channels by Palomar and Verd{\'u} in \cite{Palomar06}, can be directly applied to the model in \eqref{eqn_CS_noisy} so that:
\begin{align}
\nabla_{\M} ~ \Ixy = \SWi~\M~ \E
\label{eqn_I_MMSE}
\end{align}
where the MMSE matrix is\footnote{Note that the MMSE matrix $\E$ is a function of the kernel $\M$}:
\begin{align}
\E &= \mathbb{E} \left\{\left({\bf x} - \mathbb{E} \left\{{\bf x}|{\bf y}\right\}\right)\left({\bf x} - \mathbb{E} \left\{{\bf x}|{\bf y}\right\}\right)^\dagger  \right\} \label{eq:MMSE}\\
&= \UE ~ \LE ~ \UEt
\end{align}
where  $\UE \in \mathbb{S}^{m}$, $\LE={\sf Diag} \left( {\lambda_{E_1}}, \ldots, {\lambda_{E_m}} \right)$ and ${\lambda_E}_1 \geq {\lambda_E}_2 \geq \ldots \geq {\lambda_E}_m \geq 0$ are the (non-negative) eigenvalues of the MMSE matrix. 

Our design approach also draws on a specific kernel design constraint. It is important to recall that in CS applications the kernel design is typically set to obey unit-norm row constrains or, instead, orthonormal constraints \cite{Nenadic07}. In contrast, in communications applications the kernel (or precoder) design obeys a power (trace) constraint, which states that on average the rows have unit-norm. A paper that does consider this power constraint for CS is the work by Schnitter \cite{Schniter11}, however, this is unusual in the CS field.
We adopt this more general constraint, which, in addition to leading to solutions with higher mutual information or R{\'e}nyi entropy, enables the formulation of the design framework which the unit-norm rows constraint does not. The exception to this is the special case of adaptive online design in Section \ref{sec:online}, where each row of the kernel is designed sequentially such that the two constraints coincide.

\section{Mutual Information based Kernel Design} \label{sec_mutual_information}

In this section we consider the characterization of the kernel that maximises the mutual information of the model in \eqref{eqn_CS_noisy}, subject to a power constraint, for multivariate Gaussian sources and general multivariate sources. The optimal kernel design for multivariate Gaussian sources also provides a rationale for other kernel designs in subsequent sections, most notably, the PDS method (we extend the work of \cite{Sapiro}). The design problem can be posed abstractly as follows:
\begin{equation}
\begin{aligned}  
& \underset{ \M}{\text{maximize}}
& & \mathcal{I} \left( \x; \M ~\x + \w \right)  \\  
& \text{subject to}  
& & \frac{1}{\ell}~{\sf tr} \left( \M \Mt \right) \leq 1
\end{aligned}  
\label{eqn_opt_problem}
\end{equation}

It is important to remark that this optimization problem is non-convex in general. The use of the fundamental result in \eqref{eqn_I_MMSE}, in addition to enabling the full or partial characterization of the solution, also leads to efficient computational procedures. We restate next the characterizations of the optimal kernel designs for Gaussian sources (Theorem \ref{theorem_optimal_measurement_matrix_gaussian}) and general sources (Theorem \ref{theorem_optimal_measurement_matrix_general}), which also appear in slightly different forms in \cite{PerezCruz10} \cite{Payaro09} \cite{Lamarca09}, in a manner that emphasizes the operational significance for CS applications.

\subsection{Multivariate Gaussian Input Source} \label{section_design_gaussian_source}

The characterization of the optimal kernel design for a multivariate complex-valued Gaussian source leverages the well-known closed-form mutual information expression given by:
\begin{align}
\Ixy & = \log \det \left( \I_m +  \Mt \SWi \M ~\SX \right).
\label{eqn_Ixy_gaussian}
\end{align}
This simple closed-form expression allows the use of simple matrix identities, rather than the gradient result in \eqref{eqn_I_MMSE}, to obtain the solution to \eqref{eqn_opt_problem}. The case when $\SX=\I$ is well-known from communications theory and was recently applied in the design of measurement kernels by Schniter \cite{Schniter11}. However, the case for general source covariance matrices has not been studied in the communications domain. We unveil that this leads to the novel operation of \emph{mode alignment}\footnote{This result was also recently shown in radar \cite{Tang10}.}.

\vspace{10pt}

\begin{theorem}
\label{theorem_optimal_measurement_matrix_gaussian}
The kernel matrix that solves the optimization problem in \eqref{eqn_opt_problem} for a multivariate complex-valued Gaussian source with covariance matrix $\SX$ is given by\footnote{Note that the superscript $\star$ denotes an optimal solution.}:
\begin{align}
\Ms = \UW ~ \LMs ~\UXt
\label{eqn_optimal_measurement_matrix_gaussian}
\end{align}
where $\LMs = \left[ {\sf Diag} \left( \sqrt{\lambda^\star_{M_1}}, \ldots, \sqrt{\lambda^\star_{M_\ell}} \right) \hspace{3mm} {\bf 0}_{\ell \times (m - \ell)} \right]$, $\lambda^\star_{M_i} =
\left( \frac{1}{\eta} -  \frac{\lambda_{w_{i}}}{{\lambda_{x_i}} } \right)^+$ with the noise covariance eigenvalues ${\lambda_w}_i$ arranged in ascending order and the source covariance eigenvalues ${\lambda_x}_i$ arranged in descending order and $\eta$ ensures the \emph{average} unit-norm row constraint, i.e., $\frac{1}{\ell}~{\sf tr} \left( \M \Mt \right) = 1$.
\end{theorem}
\vspace{10pt}
\begin{proof}
See Appendix \ref{appdx_proof_gaussian}.
\end{proof}
\vspace{10pt}

Theorem \ref{theorem_optimal_measurement_matrix_gaussian} uncovers the operations of the optimal kernel design. In particular, it is possible to recognize a novel mode alignment operation which involves two aspects: i) exposing the modes of the noise and source covariance; and ii) ordering (or aligning) the modes.

\begin{figure}[!t]
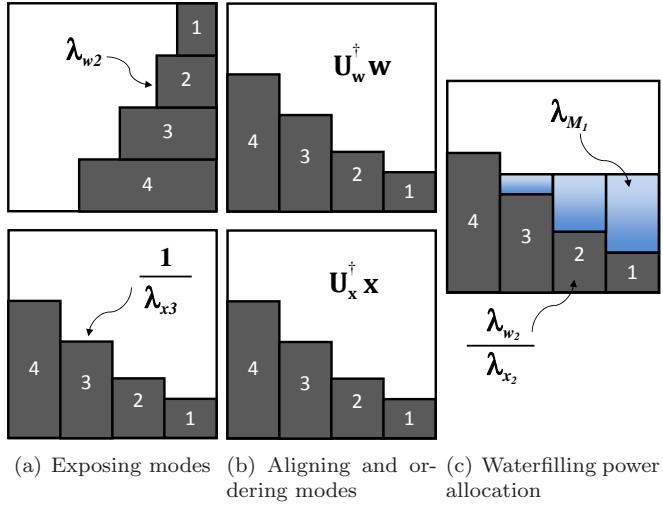

\centering
\subfigure[Exposing modes]{\label{diag_WF_modes}\includegraphics[width=28mm]{Modes_SW_SX.prn}}
\subfigure[Aligning and ordering modes]{\label{diag_WF_align}\includegraphics[width=28mm]{Align_SW_SX.prn}}
\subfigure[Waterfilling power allocation]{\label{diag_WF_WF}\includegraphics[width=28.5mm]{WF_SW_SX.prn}}
\caption{Diagrammatic view of the actions of the optimal kernel design.}
\label{diag_waterfilling}
\end{figure}

First, the left-singular vectors of the kernel are chosen to align with the eigenvectors of the noise covariance matrix and the right-singular vectors of the kernel are chosen to align with the eigenvectors of the signal covariance matrix (Fig. \ref{diag_WF_modes}). This is referred to as exposing the modes.

The ordering (or alignment) of the exposed modes is very particular, the largest source eigenvalue is matched to the smallest noise eigenvalue, the second largest source eigenvalue is matched to the second smallest noise eigenvalue, and so on (Fig. \ref{diag_WF_align}).

Finally, the kernel ``weights" the matched modes according to a ``waterfilling'' interpretation \cite{Thomas_and_Cover} (Fig. \ref{diag_WF_WF}). Intuitively, this emphasizes the less noisy ``channels" and reduces the influence of the noisier ones as a means to maximize further mutual information.

\begin{figure}[!t]
\centering
\includegraphics[width=90mm]{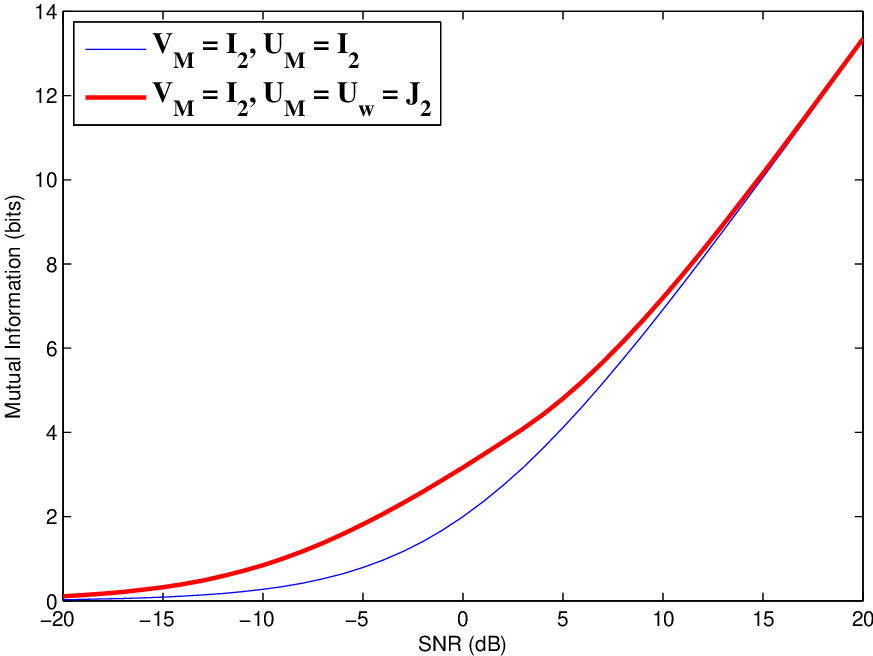}
\caption{Mutual information as a function of SNR for two different alignments both with optimal power allocation.}
\label{fig_modealignment_example2x2}
\end{figure}

As an example, Fig. \ref{fig_modealignment_example2x2} depicts the mutual information associated with two possible alignments for the signal and noise eigenvalues in a scenario where both covariance matrices are diagonal, $\LX = {\sf Diag} \left( 1, 0.25 \right)$ and $\LW = {\sf Diag} \left( 1, 0.25 \right)$. It is evident that the ordering of the modes has a significant impact on the mutual information at low and medium SNR -- the highest mutual information corresponds to the kernel design that aligns the strongest source eigenvalue with the weakest noise eigenvalue, $\UM=\UW=\J_2$ and $\VM=\I_2$.

\subsection{General Multivariate Input Source}
\label{design_general_source}

While the application of communications theory results for Gaussian distributions are known to varying degrees outside the field of communications theory, the results for general sources have not been fully leveraged outside of communications. 
The characterization of the optimal kernel design for a general multivariate complex-valued source, in view of the absence of closed-form mutual information expressions, now leverages the fundamental result in \eqref{eqn_I_MMSE}.
\vspace{10pt}

\begin{theorem}
\label{theorem_optimal_measurement_matrix_general}
The kernel matrix that solves the optimization problem in \eqref{eqn_opt_problem} for a general multivariate complex-valued source with covariance matrix $\SX$ is given by:
\begin{align}
\Ms = \UW ~  \LMs ~\VMst \label{eq:gen_theorem}
\end{align}
where $\VMs = \UEs~\Os$ \footnote{Note that the MMSE matrix $\E$ is a function of the kernel $\M$.}, the matrix $\Os$ is the optimal permutation matrix, $\LMs = \left[ {\sf Diag} \left( \sqrt{\lambda^\star_{M_1}}, \ldots, \sqrt{\lambda^\star_{M_\ell}} \right) \hspace{2mm} {\bf 0} \right]$, and $\lambda_{M_i}^\star$ are given by the generalized mercury waterfilling solution, i.e., 
\begin{align}
\lambda_{M_i}^\star = 
\begin{cases}
0, \qquad  \eta~ \lambda_{w_{i}} > {\sf mmse}_i \left( \UEs~\Os, \LMqs|_{\lambda_{M_{i}}^{\star}=0} \right)\\
\label{eqn_lambdaMi_KKT}
{\sf mmse}^{-1}_{i} \left( \eta~ \lambda_{w_{i}} \right), \qquad  \mbox{otherwise}
\end{cases}
\end{align}
where $\eta$ ensures the \emph{average} unit-norm row constraint, i.e., $\frac{1}{\ell}~\tr(\Ms \Mst) = 1$, $\LMqs = \LMs \LMst$, $\LMqs|_{\lambda_{M_{i}}^\star=0} = {\sf Diag} \left(\lambda_{M_1}^\star,\ldots,\lambda_{M_{i-1}}^\star,0,\lambda_{M_{i+1}}^\star,\ldots,\lambda_{M_\ell}^\star\right)$ and ${\sf mmse}_i\left( \UEs~\Os, \LMqs \right)$ denotes the $i$-th diagonal entry of the MMSE matrix associated with the estimate of $\x' = \Ost \UEst ~\x$ from
\begin{align}
\y' = \LWisqrt ~ \LMs~\x' + \n,
\end{align}
where $\n$ is zero-mean circularly symmetric complex Gaussian noise with identity covariance, $\n \sim \mathcal{CN}(\n; {\bf 0},\I)$. Note that ${\sf mmse}_i^{-1}$ is the inverse of ${\sf mmse}_i$ with respect to $\lambda_{Mi}$ for fixed $\lambda_{M_j}, \forall j \neq i$.
\end{theorem}

\vspace{0.25cm}

\begin{proof}
See Appendix \ref{appdx_proof_general}.
\end{proof}

\vspace{10pt}
\begin{remark}
In the high noise/low signal power regime, a first-order expansion of the mutual information is given \cite{Palomar06}:
\begin{align}
\Ixy = \frac{1}{2}~ {\sf tr} \left( \SWi~\M~\SX~\Mt \right) + o (||\SX||)
\end{align}
which implies the result observed by Shannon that at low signal-to-noise ratios proper complex discrete inputs offer a negligible loss in performance terms with regards to the capacity achieved by Gaussian inputs; hence the results for the Gaussian in Theorem \ref{theorem_optimal_measurement_matrix_gaussian} also apply in general for proper complex sources in the high noise/low power regime.
\end{remark}

\vspace{15pt}

Theorem \ref{theorem_optimal_measurement_matrix_general} suggests that the \emph{mode alignment} is no longer between the eigenvectors of the source covariance and the eigenvectors of the noise covariance, but between the eigenvectors of the MMSE matrix and the eigenvectors of the noise covariance.
The diagonalization of the MMSE matrix was first noted for communications by Lamarca \cite{Lamarca09} for identity source covariances, and the same holds true for CS for general source covariances. 
The singular values of the kernel are described by the mercury waterfilling algorithm \cite{Lozano06} \cite{PerezCruz10} which differs from waterfilling by adjusting for the non-Gaussian nature of the inputs, however, the procedure is remarkably similar.

It is important to emphasize that Theorem \ref{theorem_optimal_measurement_matrix_gaussian} characterizes fully the optimal kernel design but - in view of the non-convexity of the problem - Theorem \ref{theorem_optimal_measurement_matrix_general} characterizes partially, via a fixed point equation, the optimal kernel since $\UEs$ is still a function of $\M$. The characterization is useful because it leads to i) stopping criteria for gradient descent algorithms via \eqref{eqn_I_MMSE}; and ii) alternative optimization algorithms.
Note that if we implement gradient descent with \eqref{eqn_I_MMSE} we may get trapped in local maxima since it is known that the mutual information is not always a concave function of $\M$ \cite{Payaro09_hessian}. However, mutual information is known to be concave in the squared singular values of $\M$, for $\UM=\UW$ and fixed $\VM$. An alternative gradient descent algorithm that leads to the global maximum by avoiding local maxima switches between optimizing the singular values and the right-singular vectors of the kernel \cite{Xiao11}. 

\section{Design with R{\' e}nyi Entropy} \label{sec_Renyi_entropy}

We consider the characterization of the kernel that maximizes the output R{\'e}nyi entropy of the model in \eqref{eqn_CS_noisy}, subject to a power constraint, for multivariate Gaussian sources and multivariate Gaussian mixture sources. The design problem can then be posed as follows:
\begin{equation}
\begin{aligned}  
& \underset{ \M}{\text{maximize}}
& & \mbox{ h}_\alpha \left( \M ~\x + \w \right)  \\  
& \text{subject to}  
& & \frac{1}{\ell}~{\sf tr} \left( \M \Mt \right) \leq 1
\end{aligned}  
\label{eqn_opt_problem_Renyi}
\end{equation}
where 
\begin{equation}
\mbox{ h}_\alpha ({\bf y}) = \frac{1}{1-\alpha}~\log \int p^\alpha({\bf y})d{\bf y}.
\label{eqn_Renyi}
\end{equation}

Note that R{\'e}nyi entropy represents a generalization of Shannon entropy given by:
\begin{align}
\mbox{h}_s({\bf y}) = -\int p({\bf y}) \log p({\bf y}) d{\bf y}
\end{align}
which is the special case when $\alpha=1$.


\subsection{Multivariate Gaussian Input Source} \label{sec_Renyi_gaussian}

For multivariate Gaussian sources, both Shannon entropy and R{\' e}nyi entropy can be expressed analytically for all values of $\alpha > 0$. In particular, the two are shown to be related in the following theorem\footnote{For quadratic R{\' e}nyi entropy this result was also in Appendix A of \cite{Scardovi05}.}:
\vspace{10pt}
\begin{theorem}
\label{theorem_renyi_shannon_gaussian}

For a multivariate Gaussian input source where $\x \sim \mathcal{CN}(\x; \boldsymbol{\mu},\SX)$, the R{\' e}nyi entropy of order $\alpha >0$ and the Shannon entropy associated with the output of the model in \eqref{eqn_CS_noisy} are related as:
\begin{equation}
\mbox{ h}_\alpha ({\bf y}) = \mbox{ h}_s ({\bf y}) - \ell\left( 1 - \frac{ \log \alpha }{{\alpha - 1}} \right),
\end{equation}
where $\mbox{ h}_s ({\bf y}) = \log \left[ (2\pi e)^\ell\det \left( \SW + \M \SX \Mt \right)\right].$
\end{theorem}

\vspace{0.25cm}

\begin{proof}
See Appendix \ref{appdx_proof_renyi_gaussian}.
\end{proof}

\vspace{0.25cm}

Theorem \ref{theorem_renyi_shannon_gaussian} leads immediately to a generalization of the I-MMSE identity in \eqref{eqn_I_MMSE} for Gaussian sources:
\begin{theorem}
\label{theorem_renyi_gradient}
For Gaussian sources, the (complex) gradient with respect to the kernel of the output R{\'e}nyi entropy of order $\alpha>0$ associated with the model in \eqref{eqn_CS_noisy} obeys the relationship:
\begin{equation}
\nabla_\M \mbox{ h}_\alpha (\y) = \SWi~\M~\E.
\label{eqn_grad_renyi_gauss}
\end{equation}
\end{theorem}

\vspace{0.25cm}

Theorem \ref{theorem_renyi_gradient} unveils that the relationship between mutual information and the MMSE matrix in \eqref{eqn_I_MMSE} also holds for all values of $\alpha>0$ for the output R{\'e}nyi entropy associated with the model in \eqref{eqn_CS_noisy} for Gaussian sources. Theorem \ref{theorem_renyi_gradient} also implies that the kernel design that maximizes the R{\'e}nyi entropy subject to a power constraint also obeys the characterization in Theorem \ref{theorem_optimal_measurement_matrix_gaussian}.

%

\subsection{Multivariate Gaussian Mixture Model Input Source}\label{sec_Renyi_general}
\label{design_GMM_source}

For Gaussian Mixture Models (GMM) the signal $\x \in \mathbb{C}^{m}$ is represented by:
\begin{eqnarray}
p(\x) = \sum_{i=1}^N p(i)  ~ \mathcal{CN}(\x; \boldsymbol{\mu}_i,{\bf \Sigma}_i),\label{eq:GMM}
\end{eqnarray}
where $p(i)$ is the probability of occurrence of mixture component $i$, $\boldsymbol{\mu}_i$ and ${\bf \Sigma}_i$ correspond to the mean and covariance matrix of the $i$-th circularly symmetric complex Gaussian distribution. 
Neither the Shannon entropy, mutual information nor the MMSE matrix are known to have closed-form expressions for GMMs. R{\'e}nyi entropy and its gradient, however, admit closed-form expressions in some instances, which lend themselves more easily to optimization via gradient descent algorithms. For example, the quadratic R{\'e}nyi entropy of the noisy projection ${\bf y}$ in \eqref{eqn_CS_noisy} is given by:
\begin{align}
\mbox{ h}_2 ({\bf y}) =- \log \sum_{i=1}^N \sum_{j=1}^N p(i)~p(j) ~ \mathcal{CN}\left({\bf 0}; \boldsymbol{\mu}_{i,j},\boldsymbol{\Sigma}_{i,j} \right)
\end{align}
where:
\begin{align}
\boldsymbol{\mu}_{i,j} &= \M \left({\boldsymbol{\mu}_i} - {\boldsymbol{\mu}_j}\right)\\ 
\boldsymbol{\Sigma}_{i,j} &= \M \left({{\bf \Sigma}_i}+{{\bf \Sigma}_j}\right) \Mt + 2\SW
\end{align}

The complex gradient with respect to $\M$ of the quadratic R{\' e}nyi entropy of the noisy projection ${\bf y}$ in \eqref{eqn_CS_noisy} for the GMM is given by:
\begin{align}
\nonumber&\nabla_{\M} \mbox{ h}_2 ({\bf y}) = \\&\frac{\displaystyle -\sum_{i,j=1}^N p(i)~p(j)~ \mathcal{CN}\left({\bf 0}; \boldsymbol{\mu}_{i,j},\boldsymbol{\Sigma}_{i,j} \right)  \nabla_{\bf \M} \log \mathcal{CN}\left({\bf 0}; \boldsymbol{\mu}_{i,j},\boldsymbol{\Sigma}_{i,j} \right)} {\displaystyle \sum_{i=1}^N \sum_{j=1}^N p(i)~p(j)~ \mathcal{CN}\left({\bf 0}; \boldsymbol{\mu}_{i,j},\boldsymbol{\Sigma}_{i,j} \right)}
\label{eq:grad_ren}\end{align}
where: 
\begin{align}
\nonumber \!\!\!\nabla_{\M} \log  \mathcal{CN}_{i,j}   = &  - \boldsymbol{\Sigma}_{i,j}^{-1}~\M~\left({{\bf \Sigma}_i} + {{\bf \Sigma}_j}\right)\\
\nonumber&+   \boldsymbol{\Sigma}_{i,j}^{-1} ~\M\left({\boldsymbol{\mu}_i} - {\boldsymbol{\mu}_j}\right) \left({\boldsymbol{\mu}_i} - {\boldsymbol{\mu}_j}\right)^\dagger \\
&\times \left\{  \Mt   \boldsymbol{\Sigma}_{i,j}^{-1} {\M}\left({{\bf \Sigma}_i}+{{\bf \Sigma}_j}\right)- {\bf I}\right\}.
\end{align}
where $\times$ denotes a matrix multiplication. The proof is given in Appendix \ref{Appdx_Renyi_gradient}.

It is interesting to note that the now celebrated I-MMSE relationship in the information theory literature also applies for R{\'e}nyi entropy of order $\alpha >0$ associated with Gaussian source models. However, this relationship does not seem to carry over for the R{\'e}nyi entropy of more general source models. In fact, it can only be shown that for a general source, which obeys some additional smoothness conditions, the gradient can be expressed as follows (The proof is a modification of the result in \cite{Palomar06}): 
\begin{align}
\nabla_{\M} \mbox{ h}_\alpha ({\bf y}) &= \alpha \SWi \int \hat{p}({\bf y}) \left( {\bf y} - \M {\bf x}_y \right) {\bf x}_y^\dagger d{\bf y}
\label{eqn_grad_Renyi}
\end{align}
where the probability distribution $\hat{p}(\y) = \frac{p^\alpha(\y)}{\int  p^\alpha({\bf y}) d{\bf y}}$ and ${\bf x}_y$ is the conditional mean estimator.

It is not difficult to appreciate that the right-hand side of \eqref{eqn_grad_Renyi} is in general different from the right-hand side of \eqref{eqn_grad_renyi_gauss} (or the right-hand side of the I-MMSE relationship in \eqref{eqn_I_MMSE}) by studying the Taylor expansion of $\nabla_{\M} \mbox{ h}_2 ({\bf y})$. For the high-noise power scenario, the first term in the expressions coincide but higher order terms do not \cite{Renna12}.

\section{Application to Compressive Sensing}\label{sec_results}

\subsection{Problem setup}

\begin{figure}[!hbtp]
  \centering
  \includegraphics[width=90mm]{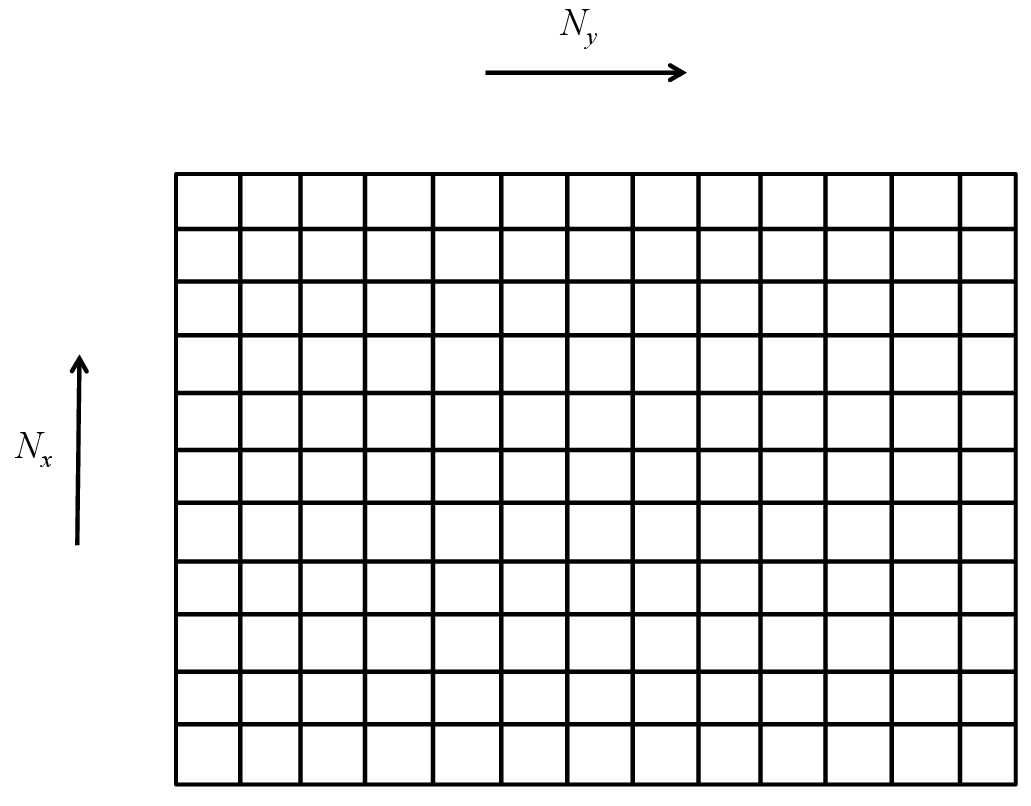}
  \caption{Spatial grid used for CS measurement image $N_x\times N_y$ image, decomposing the image into a contiguous grid of ``patches," each patch composed of $n_x\times n_y$ pixels, $n_x\ll N_x$ and $n_y\ll N_y$. Letting $\x_i\in\mathbb{R}^{n_xn_y}$ represent the pixels associated with the $i$th patch, separate projection matrices $\M_i$ are designed for each $\x_i$.}
  \label{fig:grid}
\end{figure}

We consider CS in the context of imaging. While the theory is applicable to complex data, the following examples focus on real images. Specifically, consider measurement of the image $\X\in\mathbb{R}^{N_x\times N_y}$, for large $N_x$ and $N_y$. As indicated in Figure \ref{fig:grid}, the image is partitioned into $n_x\times n_y$ contiguous ``patches,'' with the pixels in the $j$th patch denoted by vector $\x_j\in\mathbb{R}^{\ell}$, with $\ell=n_xn_y$. In the examples considered here $n_x=n_y=8$ (consistent, for example, with the patch sizes used in the JPEG standard). 

It is desirable to partition the images into such patches because one may readily learn a signal model for the $\{\x_j\}$, while it is difficult to learn an accurate signal model directly on the entire image $\X$. Specifically, following \cite{Chen10,Sapiro}, we assume that each $\x_j$ is drawn from a GMM of the form (\ref{eq:GMM}), here for \emph{real} normal distributions. 

To learn the prior signal model $p(\x)$ for the patches, we first consider a large ensemble of natural images, from which patches $\x_j\in\mathbb{R}^\ell$ are selected at random. Using these training data, a (real) GMM of the form in (\ref{eq:GMM}) is constituted as a signal model. To learn this GMM, we have employed nonparametric Bayesian methods as in \cite{Chen10}, as well as expectation-maximization (EM) methods \cite{Sapiro}, and both methods yield very similar results. The following results are based on a $N=20$ component GMM, trained on 100,000 patches, extracted at random
from $500$ natural images in the Berkeley Segmentation Dataset (\url{http://www.eecs.berkeley.edu/Research/Projects/CS/vision/grouping/resources.html}). These training images are distinct from those considered in the testing phase, for CS inversion.

While patches are selected at random from training images to constitute the prior $p(\x)$, when performing CS the goal is to recover the entire underlying image $\X$. Therefore, for CS inversion we wish to recover each of the $\{\x_j\}$ in Figure \ref{fig:grid}. In general, a separate  projection matrix $\M_j$ is applied to patch $j$ from image $\X$. For the case of offline design of the projection matrix, $\M_j$ is the same for all patches $j$ (since it is non-adaptive). For online design a distinct $\M_j$ is adaptively designed for each testing patch $j$. The measured data associated with patch $j$ is expressed as
\begin{equation} \y_j=\M_j\x_j+\w_j,~~~~~~~~j=1,\dots,J\end{equation}
In the examples that follow, the images under test are $256\times{}256$, and therefore this procedure was employed on $J=1024$ non-overlapping patches of size $8\times{}8$. Each of the $\x_j$ are recovered independently from the respective measured $\y_j$, thereby allowing for massive parallelization.

For simplicity, we henceforth drop the subscript $j$, and the discussion that follows applies to each of the $J$ patches in Figure \ref{fig:grid}. We assume the noise $\w\sim\mathcal{N}(0,\bds{\Sigma}_{\w})$, with known covariance matrix $\bds{\Sigma}_{\w}$. In the following examples we consider low-noise, i.i.d. measurements, and therefore $\bds{\Sigma}_{\w}=10^{-6}\I_\ell
$. The likelihood function for the underlying signal $\x$ is
$\mathcal{N}(\bds{y};\M \bds{x},\bds{\Sigma}_{\w})$, and the prior $p(\x)$ is the aforementioned GMM, $p(\bds{x}) = \sum_{i=1}^{N}w_{i}\mathcal{N}(\bds{x};\bds{\mu}_{i},\bds{\Sigma}_{i})$.
Under this likelihood function for $\x$, and with the GMM prior, the posterior $p(\x|\y)$ is also a GMM:
\begin{equation}\label{inference}
     p(\bds{x}|\bds{y}) = \sum\nolimits_{i=1}^{N}\widetilde{w}_{i}\mathcal{N}(\bds{x};\widetilde{\bds{\mu}}_{i},\widetilde{\bds{\Sigma}}_{i}) 
\end{equation}   
with
\begin{equation}   
\widetilde{\bds{\Sigma}}_{i}^{-1} = \bds{\M}^{T}\bds{\Sigma}_{\w}^{-1}\bds{\M}+\bds{\Sigma}_{i}^{-1},~~\widetilde{\bds{\mu}}_{i} = \widetilde{\bds{\Sigma}}_{i}(\bds{\M}^{T}\bds{\Sigma}_{\w}^{-1}\bds{y}+\bds{\Sigma}_{i}^{-1}\bds{\mu}_{i})\end{equation}
\begin{equation} \widetilde{w}_{i} = w_{i}\mathcal{N}(\bds{y};\bds{\M}\bds{\mu}_{i},\bds{\M}\bds{\Sigma}_{i}\bds{\M}^{T}+\bds{\Sigma}_{\w})/p(\bds{y})\end{equation}
\begin{equation}p(\bds{y}) = \sum\nolimits_{i=1}^{N}w_{i}\mathcal{N}(\bds{y};\bds{\M}\bds{\mu}_{i},\bds{\M}\bds{\Sigma}_{i}\bds{\M}^{T}+\bds{\Sigma}_{\w})\end{equation}
When presenting results, the estimated signal $\hat{\x}$ is the mean based on $p(\x|\y)$, $i.e.$, $\hat{\x}=\sum_{i=1}^N\widetilde{w}_i\widetilde{\bds{\mu}}_i$.
%
%
\subsection{Offline and online design\label{sec:online}}

We consider online and offline design of the projection matrix $\M$, based upon gradient descent: $\bds{\M}\leftarrow\bds{\M}+\gamma\nabla_{\bds{\M}} \mathcal{I}(\bds{x};\bds{y})$, with re-normalization to satisfy the
power constraint; here we perform a gradient of the mutual information, and the same type of gradient descent is performed in the context of R{\'e}nyi entropy, for which we therefore employ the results of Section IV. When employing the gradient of R{\'e}nyi entropy, we employ (\ref{eqn_I_MMSE}). The design of $\M$ based upon a gradient of mutual information is denoted PV, for Palomar and Ver{\'u}. 

For offline PV design, $p(\x)$ corresponds to the learned prior GMM, and the entire $\M$ is inferred at once. For online PV design, after measuring the first $k$ components of $\y$, denoted $\y_{1:k}$, we update the posterior $p(\x|\y_{1:k})$ via (\ref{inference}), and row $k+1$ of $\M$ is constituted based upon this posterior signal model; after each measurement, the posterior is updated, followed by design of the next row of $\M$, used to define the next measurement. In these computations, the MMSE matrix in (\ref{eq:MMSE}) is computed via Monte Carlo integration, based on draws from $p(\x)$ (in the offline case) or $p(\x|\y_{1:k})$ (in the online case). Online design of the patch-dependent projection matrix $\M$ may be performed in parallel.

For R{\'e}nyi-based design we consider the case $\alpha=2$; this is convenient, as within the context of the GMM representation employed here the gradient with respect to $\M$ is analytic, via (\ref{eq:grad_ren}).

The online PV design is relatively expensive, as one must repeatedly perform Monte Carlo integration to update the MMSE matrix $\E$, and one must also perform gradient descent. For online R{\'e}nyi-based design we employ (\ref{eq:grad_ren}); while this analytic expression precludes the need to numerically compute $\E$, the large number of sums makes online R{\'e}nyi and online PV design comparably expensive.

The relative expense of R{\'e}nyi and PV online design motivates a simplified online design.
In \cite{Sapiro} the authors proposed the PDS method, in which a GMM was used for $p(\x)$. In \cite{Sapiro}, the components of the first $k<\ell$ rows of $\bds{\M}$ are drawn i.i.d. from a zero-mean normal distribution. Using this $k$-row sensing matrix, an initial measurement $\y_{1:k}\in\mathbb{R}^k$ is performed. Based upon $\y_{1:k}$, the most probable mixture component from the prior $p(\x)$ is selected. At this point a single-Gaussian signal model is constituted. The remaining $\ell-k$ rows of $\M$ are then defined by the principal $\ell-k$ eigenvectors of the covariance matrix from this Gaussian. While \cite{Sapiro} did not have access to our Theorem 1, the design so constituted is consistent with it. Specifically, Theorem 1 applies to the case of a single-Gaussian signal model. Under the aforementioned assumptions for $\bds{\Sigma}_\w$ (diagonal covariance matrix, with small diagonal variance), Theorem 1 implies that the optimal projection matrix corresponds to the principal eigenvectors of the covariance matrix. However, the assumption of $k$ initial random projections employed in \cite{Sapiro}, before selecting a single Gaussian component, seems undesirable. Further, in \cite{Sapiro} the single Gaussian was selected from the prior $p(\x)$ rather from the updated posterior $p(\x|\y_{1:k})$.

We extend the PDS technique to an online setting as follows. We first initialize $p(\bds{x})$ with the GMM prior signal model (learned using offline training data). We then sequentially constitute one row of $\M$ at a time, from $k=1,\dots,\ell$; after each row is so constituted, a single new projection measurement is performed with that new row. Again let $\y_{1:k}$ represent the vector of data constituted in this manner via the first $k$ rows of $\M$. Based upon these data we update the signal model $p(\x|\y_{1:k})$. To design row $k+1$ of $\M$, let $i' =  \textrm{argmax}_{i} \widetilde{w}_{i}$, where the $\{\widetilde{w}_{i}\}$ are the GMM mixture weights from $p(\x|\y_{1:k})$. Then the $(k+1)$th row of $\M$ is defined by the leading eigenvector of $\widetilde{\bds{\Sigma}}_{i'}$.  The online PDS approximates the posterior GMM at each step with the dominant Gaussian from the posterior GMM $p(\x|\y_{1:k})$, and then via Theorem 1 the next row of $\M$ is defined by the leading eigenvector of the associated covariance matrix. Since no Monte Carlo simulation and gradient descent are needed in the above process, online PDS method is very fast. The eigenvectors are orthonormal, and therefore the power constraint is satisfied automatically at every step. Note that the posterior $p(\x|\y_{1:k})$ continuously updates with increasing data, and therefore it is not particularly sensitive to the prior $p(\x)$; the original PDS in \cite{Sapiro} was based upon the prior $p(\x)$ only, which may necessitate more care in selection of the training data. Since the posterior can be updated easily via (\ref{inference}), it appears highly preferable to use this approach rather than fixing the signal model.

\subsection{Experimental Results}
\begin{figure}[!t]
  \centering
  \includegraphics[width=90mm]{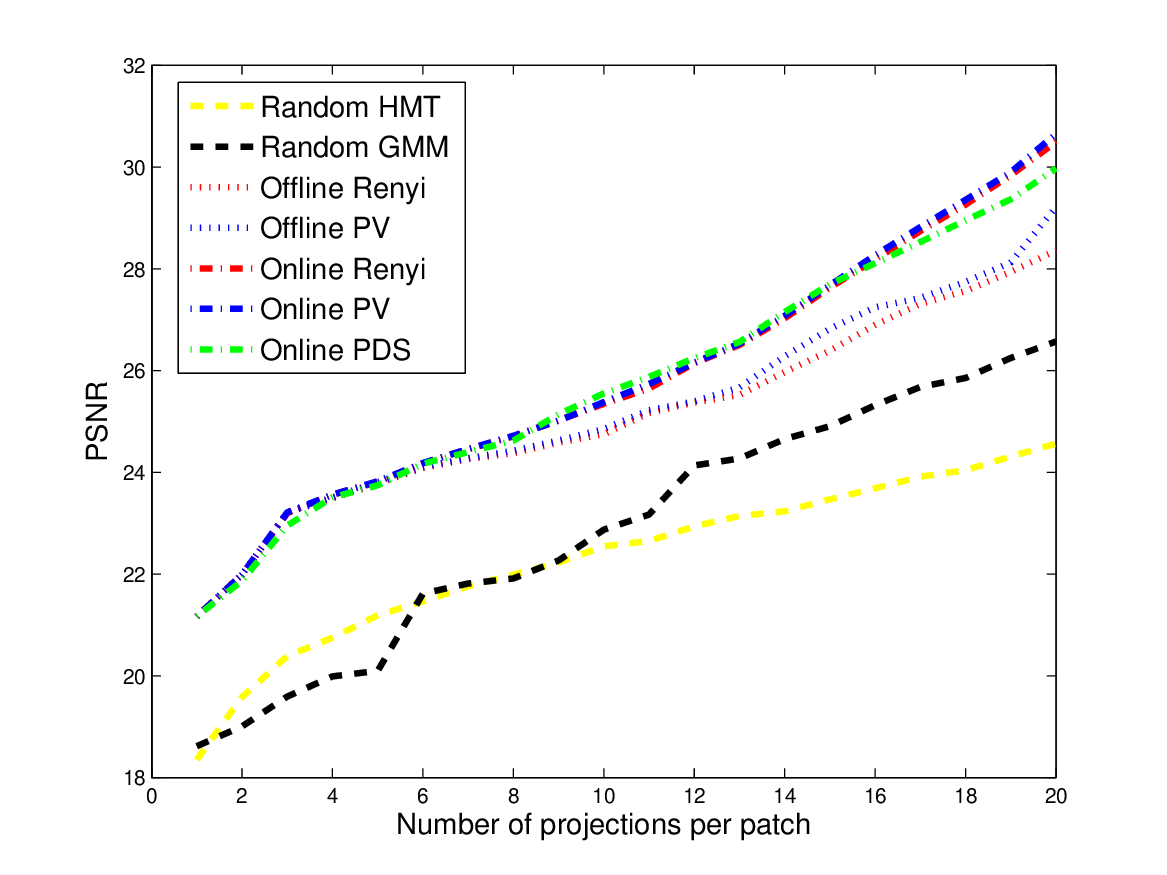}
  \caption{PSNR for the reconstructed `barbara' image.}
  \label{Barbara}
\end{figure}
\begin{figure}[!t]
  \centering
  \includegraphics[width=90mm]{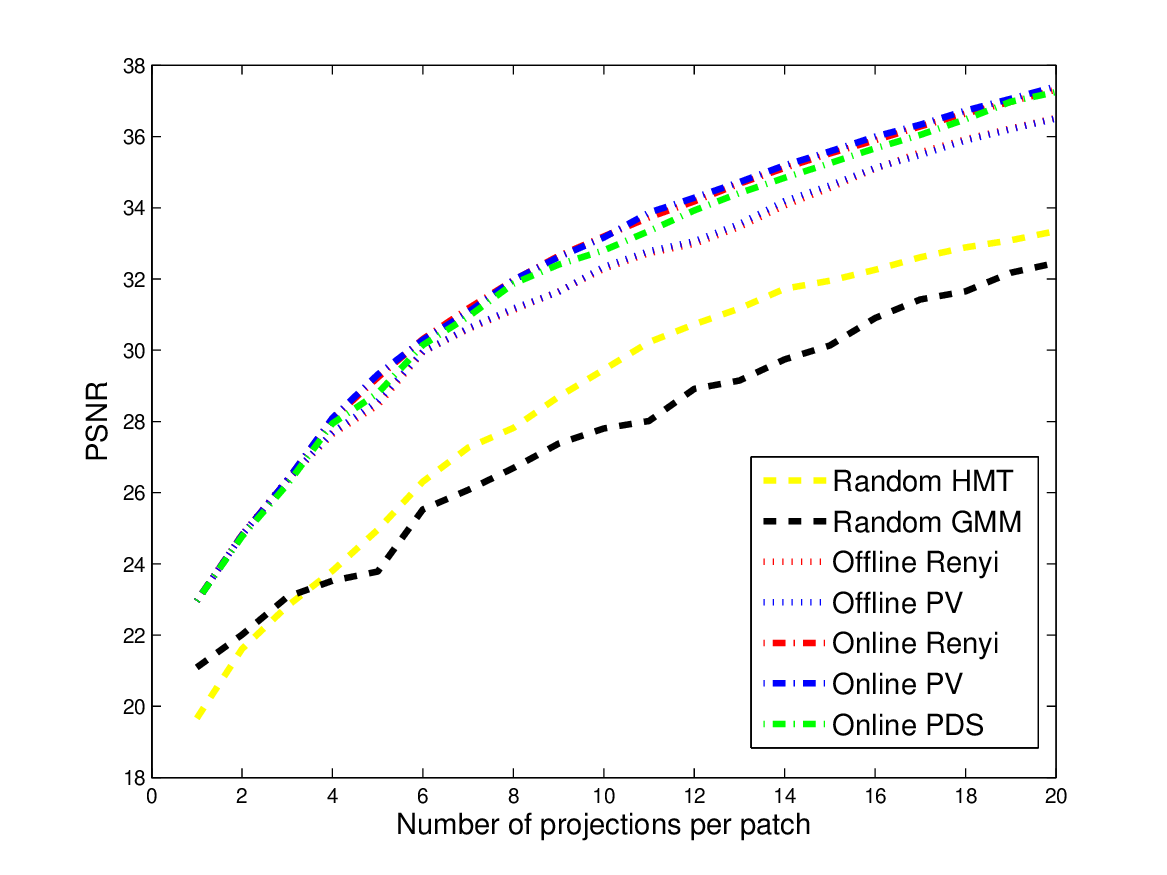}
  \caption{PSNR for the reconstructed `house' image.}
  \label{House}
\end{figure}
\begin{figure}[!tp]
  \centering
  \includegraphics[width=90mm]{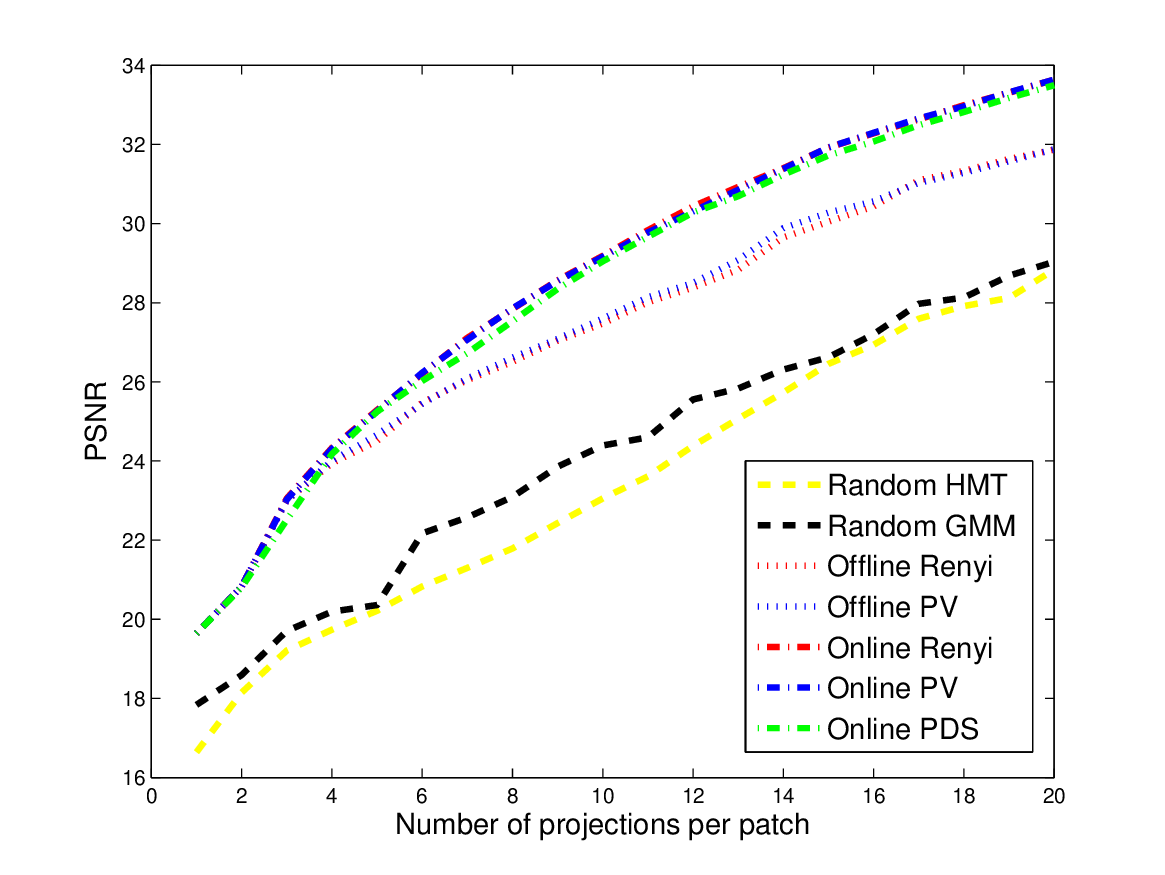}
  \caption{PSNR for the reconstructed `pepper' image.}
  \label{Pepper}
\end{figure}

In Figures \ref{Barbara}-\ref{Pepper} results are shown for three widely examined test images: `barbara', `house' and `pepper,' respectively. Two classes of results are considered based upon random projection design. The ``random GMM'' results employ the patch-based CS construction in Figure \ref{fig:grid}, and the learned GMM-based prior $p(\x)$. The form of these results are the same as employed for the designed $\M_j$, except here each $\M_j$ is constituted with matrix elements drawn i.i.d. from $\mathcal{N}(0,1)$, followed by normalization. We also considered CS design in which the projections are performed directly on the entire image $\X$, rather than at the patch level, as in Figure \ref{fig:grid}. If one performs CS inversion based on traditional CS algorithms, which employ $\ell_1$ and related regularization \cite{Candes08}, the quality of the inversion is markedly worse than that using the proposed approach, with learned signal models $p(\x)$; we therefore do not show these results here, because they don't fit on the same scale of results presented. This is not surprising, as the patch-dependent learned signal model $p(\x)$ is much richer, and tailored to the data than simple sparsity constraints, which motivate $\ell_1$ regularization. To provide a fairer comparison, when performing inversion for the case in which the projections are performed directly on the entire image $\X$, we consider an underlying wavelet basis and perform inversion based on the sophisticated hidden Markov tree (HMT) wavelet model for images \cite{Lihan}. This signal model $p(\x)$ could in principle also be used within the theory to design a projection matrix applicable to the entire image. However, the significant advantage of the GMM construction is that the posterior of the underlying signal may be constituted analytically, while for the HMT expensive computational methods are needed \cite{Lihan}. Therefore, we only show HMT inversion results when the projection matrix is constituted at random, thereby providing a comparison of inversion quality of the GMM (patch based) and the HMT (entire image), based upon random projections.

We consider offline design of the patch-based projection matrix $\M$ based upon the R{\'e}nyi measure of entropy, as well as based upon mutual information (via the PV theory). For online R{\'e}nyi and PV design, we do \emph{not} make a simplifying single-Gaussian assumption when designing each row of $\M$. By contrast the online PDS method uses the most probable Gaussian from the posterior to design the next projection at each step (this is therefore an approximation). The PDS method is very fast, while online PV is expensive, and therefore is shown principally for comparison (may not be done in practice, where online design must be fast). 

First comparing the results based on random projections, the results based upon the (learned) patched-based GMM and based on the entire-image-based HMT are comparable in reconstruction quality. Sometimes the GMM results are slightly better, and other times the HMT results are better. However, there is no comparison with respect to computation speed. The HMT results are expensive, being based upon a Gibbs sampler \cite{Lihan}. By contrast the GMM results are very fast, with the inversion analytic. The additional big advantage of the GMM representation is that it allows convenient design of patch-dependent projection matrices, which we consider next.

Each of the designed projection methods yield significant improvement relative to random, and after approximately 6 projections per patch we note that the online results are significantly better than offline design. For the first approximately 5 measurements per patch, the offline and online results are comparable; we attribute this to an inadequate number of measurements to obtain an accurate signal model, and therefore little gain manifested by adaptivity. However, after approximately 6 measurements per patch it appears that the posterior signal model becomes accurate, yielding advantages of adaptivity. Concerning online design, inversion quality based on the simple and fast online PDS performs quite competitively relative to the online R{\'e}nyi and PV design (which do not make a simplification to a single Gaussian), despite the fact that it assumes that the patch is drawn from a single Gaussian. 

\begin{figure}[!hbtp]
  \centering
  \includegraphics[width=90mm]{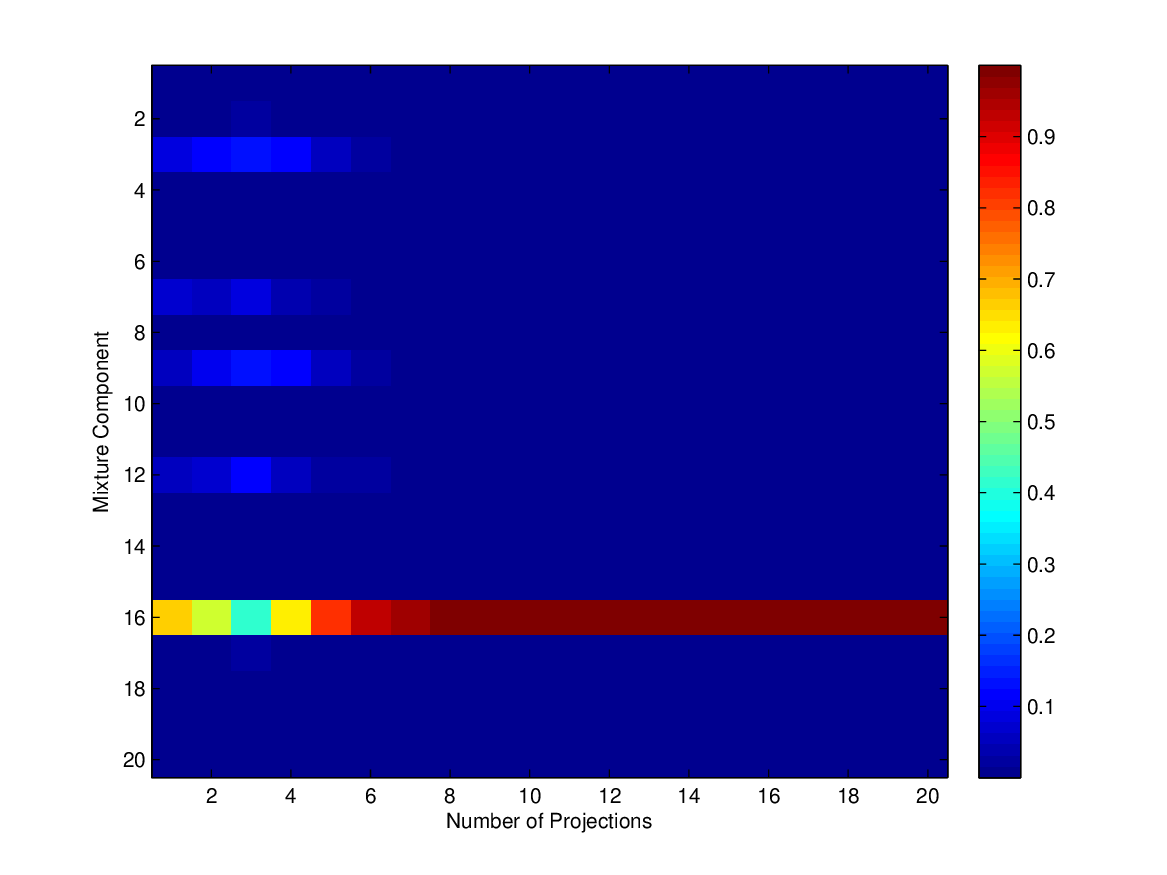}
  \caption{Evolution of the mixture weight in the posterior GMM for a typical testing patch in `barbara'.}
  \label{Weight}
\end{figure}

To understand the quality of the simple PDS-based design, consider Figure \ref{Weight}, wherein we plot the probabilities $\{\widetilde{w}_{i}\}_{i=1,N}$, for the posterior $p(\x|\y_{1:k})$, as the number of measurements $k$ increases from 1 to $\ell$. Note that after approximately six measurements the model has inferred that the underlying signal $\x$ was drawn from a single multivariate Gaussian. Note that the GMM is characteristic of an \emph{ensemble} of draws, like those characteristic of the multiple patches in Figure \ref{fig:grid}. However, any single patch is drawn from a single one of the mixture components; it is however unknown \emph{a priori} which component. Based upon experiments of this type, typically 6 projections are sufficient to infer which single mixture component a given patch corresponds to. At this point the results in Theorem 1 apply directly, which under the assumption for $\bds{\Sigma}_\w$ dictates that the optimal measurement corresponds to projecting onto the dominant eigenvector of the covariance matrix of the single mixture component (single Gaussian); this is precisely what PDS does. 

\section{Conclusions} \label{sec_conclusions}

We observe that the design principle of maximizing mutual information or R{\'e}nyi entropy leads to deterministic kernel matrices for which MMSE performance is superior to that of random kernel matrices.
In particular, we are able to provide design principles for the optimal kernel matrix for a general multivariate source that maximizes the mutual information or R{\'e}nyi entropy (for which Shannon entropy is a special case). We showed that the optimal kernel exposes the modes of the noise and the modes of the (optimal) MMSE matrix, then performs an alignment operation whose purpose it to optimally match the modes of the noise to the modes of the MMSE matrix (or, in the multivariate Gaussian source scenario, the modes of the source covariance). Finally, it carries out a generalized mercury-waterfilling power allocation operation.

The theoretical framework has been demonstrated with application to compressive sensing (CS) as applied to imagery. Using a GMM signal model, it was demonstrated that designed measurement kernels can yield markedly improved CS signal recovery relative to random design. The GMM representation has the advantage of yielding closed-form CS inversion, which is particularly attractive for fast signal inversion and for online kernel design. We have enhanced an online kernel design framework first proposed in \cite{Sapiro}, and have also provided a theoretical foundation for why it works so effectively in practice.

\bibliographystyle{siam}
\bibliography{references_all,Journal,references,BPFA_journal,MI}

\pagebreak

\appendix

\section{Complex Derivatives and Gradients}

Throughout the paper we adopt the definition of the formal partial complex derivative of a real-valued scalar function $f$ with respect to a complex-valued variable $\x$ given by \cite{Palomar06} \cite{Xiao11}:
\begin{align}
\frac{\partial f}{\partial x^*} &\triangleq \frac{1}{2} \left[ \frac{\partial f}{\partial \mbox{Re}(x)} + j \frac{\partial f}{\partial \mbox{Im}(x)} \right]
\end{align}

The definition of the complex gradient of a real-valued function $f$ with respect to a complex-valued matrix $\X$ is given by:
\begin{align}
\nabla_\X f \triangleq \frac{\partial f}{\partial \X^*}
\label{eqn_complex_grad}
\end{align}
where $\left[ \nabla_\X f \right]_{ij} = \nicefrac{\partial f}{\partial [\X^*]_{ij}}$.

\section{Helpful Lemmas}

In the proofs of the Theorems stated in this paper we will find the following lemmas helpful 

\begin{lemma}[Sylvester's Determinant Theorem] \label{lemma_cyclic_det}
We have a ``cyclic" property of determinants for two matrices ${\bf A} \in \mathcal{C}^{n \times m}$ and ${\bf B} \in \mathcal{C}^{m \times n}$:
\begin{align}
\det \left( \I_n + {\bf A}{\bf B} \right) = \det \left( \I_m + {\bf B}{\bf A} \right).
\end{align}
\end{lemma}

\vspace{0.25cm}

In the following four lemmas we denote two Hermitian matrices by ${\bf \Sigma_A}, {\bf \Sigma_B} \in \mathcal{C}^{m \times m}$ which have eigenvalues ${\lambda}_1 \geq  \cdots \geq {\lambda}_m$ and $\mu_1 \geq  \cdots \geq \mu_m$, respectively. The eigenvalue decomposition of these two matrices are ${\bf \Sigma_A} = {\bf U_A}~ {\bf \Lambda_A}~ {\bf U_A^\dagger}$ and ${\bf \Sigma_B} = {\bf U_B}~ {\bf \Lambda_B}~ {\bf U_B^\dagger}$, where ${\bf U_A},{\bf U_B} \in \mathbb{S}^{m \times m}$, ${\bf \Lambda_A}= {\sf Diag}\left({\lambda}_1, \cdots, {\lambda}_m\right)$ and ${\bf \Lambda_A}= {\sf Diag}\left({\mu}_1, \cdots, {\mu}_m\right)$.

\vspace{0.25cm}

\begin{lemma}[Theorem 1.3.12 in \cite{Horn_Johnson}] \label{lemma_sim_diag}
The matrices ${\bf \Sigma_A}$ and ${\bf \Sigma_B}$ commute if and only if they are simultaneously diagonalizable, i.e., both $\U {\bf \Sigma_A} \U^\dagger$ and $\U {\bf \Sigma_B} \U^\dagger$ are diagonal matrices for some unitary matrix $\U$.
\end{lemma}

\vspace{0.25cm}

\begin{lemma}[Richter \cite{Richter58}] \label{lemma_trace}
\begin{align}
\sum_{i=1}^m \lambda_i ~\mu_{m+1-i} \leq \tr \left( {\bf \Sigma_A}{\bf \Sigma_B} \right) \leq \sum_{i=1}^m \lambda_i~ \mu_i.
\end{align}
\end{lemma}
\begin{remark} \label{remark_richter}
Sufficient conditions for achieving the upper and lower bounds are ${\bf U_A}={\bf U_B}$ and ${\bf U_A}={\bf U_B} ~\J_m$, respectively. The sufficient condition to achieve the lower bound was given by K{\"o}se and Wesel in Theorem 2 in \cite{Kose03} and Theobald \cite{Theobald74} also gave necessary and sufficient conditions for achieving the upper bound, which allow for the multiplicity of eigenvalues.
\end{remark}

\vspace{0.25cm}

\begin{lemma}[Lemma 3 in Witzenhausen \cite{Witsenhausen75}] \label{lemma_Witsenhausen}
\begin{align}
\det \left( \I_m + {\bf \Sigma_A}{\bf \Sigma_B} \right) \leq \prod_{i=1}^m \left( 1  + \lambda_i~ \mu_i \right).
\end{align}
\end{lemma}
\vspace{0.25cm}
\begin{remark}
A sufficient condition for achieving the upper bound is ${\bf U_A}={\bf U_B}$. Witzenhausen gave further sufficient conditions which allow for the multiplicity of eigenvalues, stating that if equality holds then ${\bf \Sigma_A}$ and ${\bf \Sigma_B}$ commute and the diagonalizing matrix is such that the eigenvalues are aligned in the same order.

\end{remark}

\vspace{0.25cm}

\begin{lemma} \label{lemma_palomar}
Let $\PP \in \mathcal{C}^{m \times n}$ denote a rectangular matrix, ${\bf \Sigma_H} \in \mathcal{C}^{n \times n}$ denote a positive semi-definite matrix, and ${\PP}^\dagger {\bf \Sigma_H} {\PP} \in \mathcal{C}^{m \times m}$ be a diagonal matrix with diagonal elements in decreasing order (possibly with some zero diagonal elements). Then, there is a matrix of the form $\overline{\PP} = \VH ~[~  {\bf \Lambda}, ~{\bf 0}~ ]$ that satisfies:
\begin{align}
\overline{\PP}^{\dagger} {\bf \Sigma_H} \overline{\PP} = \alpha {\PP}^{\dagger} {\bf \Sigma_H} {\PP}\\ 
{\sf tr}( \overline{\PP} ~\overline{\PP}^{\dagger})  = {\sf tr}( {\PP} {\PP}^{\dagger})
\end{align}
where $\alpha \geq 1$, $\VH$ is a unitary matrix with columns equal to the eigenvectors of matrix $\bf \Sigma_H$ corresponding to the $\min(n,m)$ largest eigenvalues in decreasing order and $\bf \Lambda$ is square diagonal matrix of size $\min(n,m)$.
\end{lemma}
\vspace{0.25cm}
\begin{proof}
This is a modification of Lemma 3.16 in \cite{Palomar06_major}.
\end{proof}
\vspace{0.25cm}

\begin{lemma} \label{lemma_trAXBXt_conplex}
For the complex gradient defined in \eqref{eqn_complex_grad} and general matrices ${\bf A} \in \mathbb{C}^{m \times m}$, ${\bf B} \in \mathbb{C}^{n \times n}$ and ${\bf X} \in \mathbb{C}^{m \times n}$, we have:
\begin{align}
\nabla_\X {\sf tr}({\bf A}~{\bf X}~{\bf B}~{\bf X}^\dagger) = {\bf A}~{\bf X}~{\bf B}
\end{align}
\end{lemma}
\vspace{0.25cm}
\begin{proof}
Using properties of differentials (32) and (33) from \cite{Cookbook_2008} we have:
\begin{align}
\partial{\sf tr}({\bf A}~{\bf X}~{\bf B}~{\bf X}^\dagger)  = {\sf tr}[{\bf A}~\partial({\bf X})~{\bf B}~{\bf X}^\dagger] + {\sf tr}[{\bf A}~{\bf X}~{\bf B}~\partial({\bf X}^\dagger)].
\end{align}

Together with the results for complex derivatives (219), (220), (221) and (222) from \cite{Cookbook_2008} we have:
\begin{align}
\frac{\partial}{\partial \mathtt{Re}(\X)}{\sf tr}({\bf A}~{\bf X}~{\bf B}~{\bf X}^\dagger)  &= {\bf A}^\intercal~{\bf X}^*~{\bf B}^\intercal + {\bf A}~{\bf X}~{\bf B}\\
i\frac{\partial}{\partial \mathtt{Im}(\X)}{\sf tr}({\bf A}~{\bf X}~{\bf B}~{\bf X}^\dagger)  &= -{\bf A}^\intercal~{\bf X}^*~{\bf B}^\intercal + {\bf A}~{\bf X}~{\bf B}
\end{align}
and the result follows.
\end{proof}
\vspace{0.25cm}
\begin{remark}
This is the counterpart for complex-valued matrices to result (108) in \cite{Cookbook_2008} for real-valued matrices; note that the term $ {\bf A}^\intercal~{\bf X}~{\bf B}^\intercal$ is absent in the complex case.
\end{remark}

\vspace{0.25cm}

\begin{lemma} \label{lemma_complex_116_cookbook}
For the complex gradient defined in \eqref{eqn_complex_grad} and general matrices ${\bf A} \in \mathbb{C}^{m \times m}$, ${\bf B} \in \mathbb{C}^{n \times n}$, ${\bf C} \in \mathbb{C}^{n \times n}$ and ${\bf X} \in \mathbb{C}^{m \times n}$, we have:
\begin{align}
\nonumber&\nabla_\X {\sf tr}\left[ ( {\bf A} + {\bf X} ~ {\bf B} ~ { \bf X} ^\dagger )^{-1}~({\bf X}~{\bf C}~{\bf X}^\dagger) \right] \\
&= ({\bf A} + {\bf X} {\bf B} { \bf X} ^\dagger )^{-1}~{\bf X}~{\bf C} \left[\I - {\bf X}^\dagger~({\bf A} + {\bf X}  {\bf B}  { \bf X} ^\dagger )^{-1}~{\bf X} {\bf B}\right]
\end{align}
\end{lemma}
\vspace{0.25cm}
\begin{proof}
Using properties of differentials (32), (33) and (36) in \cite{Cookbook_2008} and the abbreviation ${\bf Y} =  {\bf A} + {\bf X} ~ {\bf B} ~ { \bf X} ^\dagger$, we have:
\begin{align}
\nonumber&\partial{\sf tr}\left[ {\bf Y}^{-1}~({\bf X}~{\bf C}~{\bf X}^\dagger) \right] =
{\sf tr} \left\{{\bf Y}^{-1}~\partial({\bf X}~{\bf C}~{\bf X}^\dagger)\right\}
\\
\nonumber& + {\sf tr}  \left\{-{\bf Y}^{-1} \partial( {\bf A} + {\bf X} ~ {\bf B} ~ { \bf X} ^\dagger ){\bf Y}^{-1}~({\bf X}~{\bf C}~{\bf X}^\dagger)\right\} 
\end{align}

Applying Lemma \ref{lemma_trAXBXt_conplex} the result follows.
\end{proof}
\vspace{0.25cm}
\begin{remark}
This is the counterpart for complex-valued matrices to result (116) in \cite{Cookbook_2008} for real-valued matrices; note that we do not require the assumption that ${\bf B}$ and ${\bf C}$ are Hermitian (symmetric) and there is no factor of $2$.
\end{remark}
\vspace{0.25cm}

\section{Proof of Theorem \ref{theorem_optimal_measurement_matrix_gaussian}}
\label{appdx_proof_gaussian}

\begin{proof}  We first provided an alternative derivation of this proof in \cite{Carson12}. The current proof is derived directly from the proof for an identical theorem for radar in \cite{Tang10}.
We restate the mutual information between the input and output of the compressive sensing model in \eqref{eqn_CS_noisy} for a multivariate Gaussian source as follows:
\begin{align}
\tag{\ref{eqn_Ixy_gaussian}}\Ixy & =  \log \det \left( \I_m + \Mt \SWi \M ~\SX  \right).
\end{align}

Note that for a unitary matrix $\U$, the kernel $\PP=\M\U$ has the same power as $\M$, i.e., $\tr (\PP \PPt) = \tr (\M \Mt)$, but it may have different mutual information. In particular, a choice of $\U$ that maximizes the mutual information for a given $\M$ is $\U=\UX$. This can be seen from Lemma \ref{lemma_Witsenhausen} and Remark \ref{lemma_Witsenhausen}.
From Lemma \ref{lemma_palomar} we know that there exists a matrix $\overline{\PP} = \UW ~[~  {\bf \Lambda}, ~{\bf 0}~ ] $, which satisfies ${\sf tr}( \overline{\PP} ~\overline{\PP}^{\dagger})  = {\sf tr}( {\PP} {\PP}^{\dagger}) $ and $ \overline{\PP}^{\dagger} {\bf \Sigma_H} \overline{\PP} = \alpha ~\PPt \SWi \PP$ where $\alpha \geq 1$.
Since the function $\det(\I + \alpha \A) $ is monotonically increasing in $\alpha$ for a positive semi-definite matrix $\A$, the optimal kernel matrix must have the form of $\Ms= \UMs \LMs \VMst = \UW ~[~  {\bf \Lambda}, ~{\bf 0}~ ]  \UXt$.

Finally, we determine the optimal singular values by optimizing the mutual information with respect to the eigenvalues rather than the singular values, since 1) they map one-to-one (up to a factor of $\exp j\theta$, which does not affect the mutual information) and 2) this new optimization problem is convex, so the Karush-Kuhn-Tucker (KKT) optimality conditions \cite{Boyd04} define the unique global optimum. This is given by:
\begin{align}
\lambda_{M_i}^\star = 
\begin{cases}
0, & \frac{1}{\eta} - \frac{ \lambda_{w_{i}}} {\lambda_{x_i}} \leq 0 , \\
\frac{1}{\eta} - \frac{ \lambda_{w_{i}}} {\lambda_{x_i}}, & \frac{1}{\eta} - \frac{ \lambda_{w_{i}}} {\lambda_{x_i}} > 0 
\end{cases}
\end{align}
where $\eta$ is such that the average unit norm row constraint is satisfied, i.e., $\frac{1}{\ell}\sum \lambda_{M_i}^\star  = 1$, where the eigenvalues of  $\SX$ are arranged in descending order and the eigenvalues of $\SW$ are arranged in ascending order , i.e., ${\lambda_x}_1 \geq  \cdots \geq {\lambda_x}_m \geq 0$ and $0 \leq {\lambda_{w1}}  \leq \cdots \leq {\lambda_{w_\ell}}$.
\end{proof}

\section{Proof of Theorem \ref{theorem_optimal_measurement_matrix_general}}
\label{appdx_proof_general}

\begin{proof} The proof draws on the work by Payar{\'o} and Palomar \cite{Payaro09}, which described the generalized mercury waterfilling aspect of the solution, but not the mode alignment aspect, and the work by Lamarca \cite{Lamarca09}, which described the mode alignment aspect but did not focus on the generalized mercury waterfilling interpretation. The current proof highlights both the mode alignment and mercury waterfilling aspects of the solution.

The solution to the optimization problem in \eqref{eqn_opt_problem} satisfies the KKT optimality conditions:
\begin{align}
\nabla_{_\M} \left\{ - \Ixy  - \eta \cdot \bigg[ \ell - {\sf tr} \left( \M \M^\dagger \right)\bigg] \right\} \bigg|_{ \M = \M^\star} \!\!= 0 \\
 \eta \cdot \bigg[ \ell - {\sf tr} \left( \Ms \Mst \right)\bigg] = 0
\end{align}
with $\eta \geq 0$. Using the relationship between the gradient of the mutual information and the MMSE matrix in \eqref{eqn_I_MMSE}, the optimal kernel satisfies:
\begin{align}
\eta \cdot \M^\star \M^{\star\dagger} = \SWi ~ \left( \Ms~\Es~\Mst  \right).
\label{eqn_MMt_Hermitian_KKT}
\end{align}
We note that \eqref{eqn_MMt_Hermitian_KKT} is diagonalized by $\UMs$, by definition, from which it can be seen that the matrices $\SWi$ and $\Ms \Es \Mst$ commute. From this observation, together with the fact that \eqref{eqn_MMt_Hermitian_KKT} is Hermitian and Lemma \ref{lemma_sim_diag}, we deduce that:
\begin{align}
{{\bf U}_{\bf M}^\star} &= \UW   {{\bf \Pi}_{\bf U}^\star} {{\bf \Lambda}_{\bf U}}
\end{align}
where ${{\bf \Lambda}_{\bf U}}$ is a diagonal matrix with unit modulus diagonal elements and ${\bf \Pi}_{\bf U}^\star$ is a permutation matrix. Furthermore, $\UMst \Ms \Es \Mst \UMs$ is a diagonal matrix, from which we can infer:
\begin{align}
{{\bf V}_{\bf M}^\star} &= \UEs {{\bf \Pi}_{\bf V}^\star} {{\bf \Lambda}_{\bf V}}
\end{align}
where ${{\bf \Lambda}_{\bf V}}$ is a diagonal matrix with unit modulus diagonal elements and ${\bf \Pi}_{\bf V}^\star$ is a permutation matrix.
Both mutual information and the MMSE matrix are independent of ${{\bf \Lambda}_{\bf U}}$ and ${{\bf \Lambda}_{\bf V}}$, allowing us to write without loss of generality the optimal unitary matrices as follows:
\begin{align}
\UMs &= \UW \label{unitary_matrix1} \\
{{\bf V}_{\bf M}^\star} &= \UEs  ~\Os \label{unitary_matrix2}
\end{align}
where $\Os$ is some optimal permutation matrix.

By setting $\UMs = \UW$ we can now obtain an equivalent\footnote{The equivalence is in the sense that the mutual information between the input and the output of both models is equal, i.e., $\Ixy = \mathcal{ I} \left( \x; \y' \right) $.} channel model:
\begin{align}
\label{model2}
\y' &= \LWisqrt ~ \LM ~ \VMt ~ \x + \n
\end{align}
where $\y' =   \LWisqrt~\UWt~ \y$ and $\n = \LWisqrt~\UWt~ \w$ is zero-mean circularly symmetric complex Gaussian noise with identity covariance, $\n \sim \mathcal{CN}(\n; {\bf 0},\I)$. 

It was shown in \cite{Xiao11} that for a fixed value of $\VM$ the mutual information $\mathcal{I}(\x; \y')$ is concave with respect to the squared singular values of $\LM$, i.e., the following optimization problem has a unique global optimum given by the KKT conditions, where $\LMq=\LM ~\LMt$:
\begin{equation}
\begin{aligned}  
& \underset{{\lambda_{M_1},\lambda_{M_2},\ldots,\lambda_{M_\ell}} }{\text{maximize}}
& & \mathcal{ I} \left( \x; \y' \right)  \\  
& \text{subject to}   & & \sum_{i=1}^{\ell} \lambda_{M_i}  \leq \ell \\
& & & \lambda_{M_i} \geq 0
\end{aligned}  
\label{eqn_new_opt_problem}
\end{equation}
The Lagrangian for this optimization problem is:
\begin{align}
\mathcal{L}(\LMq) = \mathcal{I} ( \x; \y' ) + \eta & \left( \ell - \sum_{i=1}^{\ell} \lambda_{M_i}\right) + \sum_{i=1}^{\ell} \eta_i \lambda_{M_i}
\end{align}
and the Karush-Kuhn-Tucker conditions state that:
\begin{align}
\frac{\partial}{\partial \LMq} \mathcal{L}(\LMq) \bigg|_{\LMq = \LMqs} &= 0
\label{eqn_KKT_general}
\\
\eta \cdot \left(\ell - \sum_{i=1}^{\ell} \lambda_{M_i}^\star \right) &= 0 \\
\eta_i \cdot \lambda_{M_i}^\star &= 0, \qquad i = 1,\ldots,\ell
\end{align}

By using the result from \cite{Xiao11} that states that: 
\begin{align}
\frac{\partial}{\partial \LMq } \mathcal{I} \left( \x ; \y' \right)  = {\sf Diag} \left( \E_{\x'}~ \LWi ~ \right)
\end{align}
where $\E'= \VMt~\E~\VM$ is the MMSE matrix associated with the estimation of $\x' = \VMt~\x$ from $\y'$, it is possible to rewrite \eqref{eqn_KKT_general} as follows:
\begin{align}
\eta {\lambda}_{w_i} - {\sf mmse}_i \left(\VM, \LMqs \right) = \eta_i {\lambda}_{w_i} , \; \; \qquad i = 1,\ldots,\ell.
\end{align}
where ${\sf mmse}_{i} \left( \VM, \LMq \right)$ denotes the $i$-th diagonal entry of $\E_{\x'}$ for that particular $\LMq$.


From the KKT conditions, we know that if $\lambda_{M_i}^\star > 0$ then $\eta_i=0$ and that $\eta >0$. For a given value of $\eta$, the value of $\lambda_{M_i}^\star$ can be calculated from the relationship $\eta ~\lambda_{w_{i}} = {\sf mmse}_i \left( \VM, \LMqs \right)$ for fixed $\lambda_{M_j}, \forall j \neq i$.
The function ${\sf mmse}_i$ is non-negative and monotonically decreasing in $\lambda_{M_i} \in \left[0,\infty\right]$ for fixed $\lambda_{M_j}, \forall j \neq i$, and its maximum value is given when $\lambda_{M_i} = 0$. Therefore if $\eta~ \lambda_{w_{i}} > {\sf mmse}_i \left( \VM, \LMqs|_{\lambda_{M_{i}}^\star=0} \right)$ where $\LMqs|_{\lambda_{M_{i}}^\star=0} = {\sf Diag} \left(\lambda_{M_1}^\star,\ldots,\lambda_{M_{i-1}}^\star,0,\lambda_{M_{i+1}}^\star,\ldots,\lambda_{M_\ell}^\star\right)$, then  $\lambda_{M_i}^\star = 0$ and $\eta_i \neq 0$.


This result is true for all values of $\VM$, therefore it is also true when $\VM = \UEs~\Os$ and the result follows.
\end{proof}

%
%
%

\section{Proof of Theorem \ref{theorem_renyi_shannon_gaussian}}
\label{appdx_proof_renyi_gaussian}

\begin{proof}
Note that:
\begin{align}
p({\bf y}) =& \mathcal{CN}\left({\bf y}; \M~\x,\SW + \M~\SX~\Mt\right)
\end{align}
and so:
\begin{align}
\!\!\!p^\alpha({\bf y}) =& \frac{\mathcal{CN}\left({\bf y}; \M~\x,\frac{1}{\alpha} (\SW + \M~\SX~\Mt)\right)}{{\alpha}^{k}\left(2\pi\right)^{k(\alpha - 1)} \det \left( \SW + \M~\SX~\Mt \right)^{\alpha-1}}
\end{align}

By substituting this into the expression for R{\' e}nyi entropy it follows that:
\begin{align}
\mbox{ h}_\alpha ({\bf y}) 
= \log \left[\left(2\pi\right)^k \det (\SW + \M~\SX~\Mt)\right] + \frac{k\log \alpha}{(\alpha - 1)} .
\end{align}

The result now follows using the definition of Shannon entropy for Gaussian sources.
\end{proof}

\section{Proofs for Gradient of R{\' e}nyi entropy}
\label{Appdx_Renyi_gradient}

\begin{proof}
Let us first show that we can express the following relevant gradient analytically:
\begin{align}
\nonumber & \nabla_{\M} \log \mathcal{CN}\left({\bf 0}; \boldsymbol{\mu}_{i,j}, \boldsymbol{\Sigma}_{i,j} \right) = \\& -\nabla_{\M} \bigg(k \log 2\pi \bigg) - \nabla_{\M}\left\{\log \det  \boldsymbol{\Sigma}_{i,j} \right\} \\
&- \nabla_{\M}  \left\{ \boldsymbol{\mu}_{i,j}^\intercal \boldsymbol{\Sigma}_{i,j}^{-1}  \boldsymbol{\mu}_{i,j} \right\},
\end{align}
where $\boldsymbol{\mu}_{i,j} = \M \left({\boldsymbol{\mu}_i} - {\boldsymbol{\mu}_j}\right)$ and $\boldsymbol{\Sigma}_{i,j} =\M \left({{\bf \Sigma}_i}+{{\bf \Sigma}_j}\right) \Mt + 2\SW$.

The first term is zero, the second term is the mutual information for a complex Gaussian distribution and can be evaluated using \eqref{eqn_I_MMSE}, relating the mutual information and MMSE matrix: 
\begin{align}
\nabla_{\M} \log \det  \boldsymbol{\Sigma}_{i,j} &= (2\SW)^{-1}\M ~\E_{i,j}
\end{align}
where the $\E_{i,j}=\left[\left({{\bf \Sigma}_i} + {{\bf \Sigma}_j}\right)^{-1} +   \Mt (2\SW)^{-1}  \M\right]^{-1}$ is the MMSE matrix if the input signal $\x$ was Gaussian distributed with covariance $\left({{\bf \Sigma}_i} + {{\bf \Sigma}_j}\right)$ and distorted by Gaussian noise with covariance $2\SW$. It can also be expressed:
\begin{align}
\nabla_{\M} \log \det  \boldsymbol{\Sigma}_{i,j}  =  \boldsymbol{\Sigma}_{i,j}^{-1}~\M\left({{\bf \Sigma}_i} + {{\bf \Sigma}_j}\right)
\end{align}
where we can use Woodbury's Inversion Lemma to convert between the two. 
The third and final term, using Lemma \ref{lemma_complex_116_cookbook} and chain rule (94) in \cite{Palomar06}, can be expressed:
\begin{align}
&\nonumber \nabla_{\M}  \left\{ \boldsymbol{\mu}_{i,j}^\intercal \boldsymbol{\Sigma}_{i,j}^{-1}   \boldsymbol{\mu}_{i,j} \right\} =~\boldsymbol{\Sigma}_{i,j}^{-1} ~\M\left({\boldsymbol{\mu}_i} - {\boldsymbol{\mu}_j}\right) \left({\boldsymbol{\mu}_i} - {\boldsymbol{\mu}_j}\right)^\dagger \\
&\times \left\{  \I - \Mt   \boldsymbol{\Sigma}_{i,j}^{-1} {\M}\left({{\bf \Sigma}_i}+{{\bf \Sigma}_j}\right)\right\}.
\end{align}

\end{proof}

\end{document}